\documentclass[aip,reprint]{revtex4-1}

\usepackage{graphicx}
\usepackage{float}

\begin{document}

\title{Stochastic Time-Series Spectroscopy}

\author{John Scoville}
\affiliation{San Jose State University, Dept. of Physics, San Jose, CA 95192-0106, USA}

\begin{abstract}
Spectroscopically measuring low levels of non-equilibrium phenomena (e.g. emission in the presence of a large thermal background) can be problematic due to an unfavorable signal-to-noise ratio. An approach is presented to use time-series spectroscopy to separate non-equilibrium quantities from slowly varying equilibria.  A stochastic process associated with the non-equilibrium part of the spectrum is characterized in terms of its central moments or cumulants, which may vary over time. This parameterization encodes information about the non-equilibrium behavior of the system.

Stochastic time-series spectroscopy (STSS) can be implemented at very little expense in many settings since a series of scans are typically recorded in order to generate a low-noise averaged spectrum.  Higher moments or cumulants may be readily calculated from this series, enabling the observation of quantities that would be difficult or impossible to determine from an average spectrum or from prinicipal components analysis (PCA).  

This method is more scalable than PCA, having linear time complexity, yet it can produce comparable or superior results, as shown in example applications.  One example compares an STSS-derived CO$_2$ bending mode to a standard reference spectrum and the result of PCA.  A second example shows that STSS can reveal conditions of stress in rocks, a scenario where traditional methods such as PCA are inadequate.  This allows spectral lines and non-equilibrium behavior to be precisely resolved.  A relationship between 2nd order STSS and a time-varying form of PCA is considered.  Although the possible applications of STSS have not been fully explored, it promises to reveal information that previously could not be easily measured, possibly enabling new domains of spectroscopy and remote sensing.
\end{abstract}

\maketitle
\section{Introduction}

Statistical systems tend to fluctuate around equilibria, and the non-equilibrium behavior of a system can be characterized by its fluctuations around equilibrium.  Moreover, many quantum mechanical effects, e.g. emission from radiative transitions, introduce an intrinsically random component to spectral data.  In contrast to the classical equilibrium (which is time-independent by definition) these transient random phenomena exhibit time-dependent behavior.  Repeated measurements of these random events will reveal a characteristic statistical distribution that can potentially change over time.  The associated random variables are often interpreted as noise that has been superimposed onto an equilbrium spectra, and multiple spectra are often averaged in an attempt to cancel their contribution, i.e. produce a less noisy spectrum.

We propose that, in addition to the average that is traditionally extracted from the distribution of values present in a series of spectra, higher moments of this distribution also contain useful information.  In many cases, these higher moments reveal salient spectral features (e.g. emission lines) in much greater detail than averaged spectra.  By considering the time series associated with a particular spectral band as samples from a statistical distribution, we can extract information about the non-equilbrium behavior of a system.

In this paper, we focus on series of spectra that are parameterized by time, characterizing statistical variations over this parameter.  However, the same analysis could be applied to other parameters, spatial or otherwise, or even multidimensional parameter spaces if the appropriate multivariate statistics are considered.

\section{Characterizing departures from equilibrium}

\subsection{Moments}

In order to analyze the statistical distribution of spectral fluctuations from equilibrium, it is useful to consider a parameterization of its probability density function in terms of moments, cumulants, or other parameters.  The distribution of a random variable $X$ gleaned from spectral data will generally be unknown, but it may be characterized by the moments of the distribution of $X$.  The nth raw moment is the expected value of $X^n$, where n is an integer, and may be directly estimated from sample data.  Without taking into account sample bias, this is:
\begin{equation}
m_n=E(X^n)
\end{equation}  

A moment generating function may be written formally as the expected value of $e^{tX}$:
\begin{equation}
M_n(t)= E(e^{tX})
\end{equation}

The moments $m_n$ are the coefficients in the MacLaurin expansion of $M(t)$
\begin{equation}
m_n = \frac{d^n}{dt^n} M(0)
\end{equation}

Given this series expansion, knowlege of the moments is equivalent to knowledge of the distribution of $X$.  

Sometimes 'moment' is used somewhat colloquially to refer to either the raw moment, as defined above, or the central moments.  The raw moments of $X$ are taken about $t=0$.  In many cases, however, the quantities of interest are the central moments, which are the moments taken about $t=\mu$.  These correspond to the raw moments of a distribution in which the mean has been shifted to zero.  The nth central moment is equal to the nth raw moment of the $X-\mu$.  

Whereas the first raw moment of $X$ is its mean, $\mu$, the first central moment is zero, since $X-\mu$ necessarily has mean zero.  Higher central moments correspond to familiar statistics: the second central moment is the variance, whose square root is the standard.  The third central moment the skew, and the fourth the kurtosis.  

If a set of spectra is parameterized in more than one dimension (for example, two spatial dimensions) the multivariate moments over these parameters are tensors.  A multivariate second moment, for instance, may be represented by a matrix of covariances.

\subsection{Cumulants}

Equivalently, a statistical distribution can be represented as a sequence of cumulants rather than a series of moments.  A cumulant generating function may be written as the logarithm of the moment generating function.  
\begin{equation}
g(t)= \ln E(e^{tX})
\end{equation}

The cumulants $\kappa_n$ are the coefficients in the MacLaurin expansion of $g(t)$
\begin{equation}
\kappa_n = \frac{d^n}{dt^n} g(0)
\end{equation}

Thus, the first n cumulants are equivalent to the first n moments.  The first, second, and third cumulants agreed with the first, second, and third central moments, respectively.  However, higher cumulants are polynomial functions of the higher moments.  The first cumulant is the mean $\mu_t(\nu)$, which is equal to the first raw moment.  The second cumulant, the variance, is equal to the second central moment, and the third cumulant is equal to the third central moment, the skewness.  The fourth and higher cumulants are not generally equal to the fourth and higher moments, however.  The fourth cumulant is the excess kurtosis, which is sometimes simply called kurtosis, even though it is not equal to the third central moment (kurtosis), which sometimes leads to confusion.  The fourth cumulant (excess kurtosis) is equal to the fourth central moment (kurtosis) minus three times the second central moment.  In general, central moments and cumulants beyond the third must be considered separately.  Only the delta distribution $\delta(x) = \frac{1}{2\pi} \int_{-\infty}^{\infty} e^{itx} dt$ and the normal distribution have a finite number of nonzero cumulants.

Although moments are more easily defined and calculated than cumulants, the cumulants may be more convenient than moments in certain applications.  Because the cumulants are additive, they often arise naturally in the context of statistical physics, where extrinsic quantities are often the sum of a large number of individual units.  For this reason, the higher cumulants often have a more direct physical interpretation than the higher moments.

\section{Limiting cases and related concepts}

\subsection{Relation to p-norms}

In some cases, it is useful to consider the trend of the nth moment as n becomes large.  Since the value of the nth moment tends to scale as $x^n$, it is helpful in such cases to consider the nth root of the nth moment, which we will call $L_n$, in order to maintain linearity.
\begin{equation}
L_n(X)=\sum(|X|^n)^\frac{1}{n}
\end{equation}  

Defined in this way, $L_n$ is equal to the p-norm of a vector whose components are the elements of $X$, where $n=p$.  Since the sum differs from the expectation operator by only a constant factor equal to the number of elements in the sum, the p-norm differs from the pth root of the pth moment by the pth root of this constant.  The familiar standard deviation of a Gaussian distribution, being the square root of variance, is essentially an $L_2$ norm.  One advantage of considering the p-norms rather than moments is that the p-norms are defined such that they remain linear in the units associated with the random variable $X$.

The limit as $n \rightarrow \infty$ is a special case, corresponding to the p-norm where $\lim_{p \to \infty}$, the $L_{\infty}$ norm, also called the uniform norm, Chebyshev norm, or supremum (sup) norm.  $L_{\infty}(X)$ is equal to the maximum absolute value of an element of X divided by c, the number of elements in $X$:
\begin{equation}
L_\infty=\lim_{n \to \infty} L_{n} = max(|X|)
\end{equation}

This can be very useful analyzing spectral timeseries and may be quickly evaluated without floating-point operations.  Another useful p-norm is $L_1$.  This is sometimes called the Manhattan distance or Taxicab distance.  Although this does not explicitly correspond to a moment or cumulant of the distribution, it can be a very useful way to characterize a distribution.

\subsection{Fractional p-norms and moments}

Given the definition of the nth moment, it is natural to consider cases in which n could take any real value, rather than an integer.  For non-integer values of n and p, we obtain fractional (or fractal, for short) moments and norms.  In such a case, the moments or norms become complex numbers, and there is no \emph{a priori} reason to restrict n to the real line - n could take any value in the complex plane.  Likewise, p-norms may be analytically continued to the complex plane in a straightforward manner, while also remaining linear in the original units.

This notion of fractional moments can be formalized using generating functions that are defined with fractional derivatives \cite{Dremin}, and fractional cumulants may also be defined in this way.

At non-integer values of n or p, fractional moments or p-norms oscillate due to a nonzero imaginary part.  As n becomes large, the amplitude of the complex oscillations decrease, and the fractional moments converge to their values at integers.  Thus, the benefit of fractional moments is most apparent at low values of n.  Special cases arise when n or p are in the unbounded neighborhoods of 0 and 1.  For values of p between 0 and 1, the resulting functional is technically a quasi-norm rather than a norm\cite{}, since the triangle equality is satisfied only up to a constant, but such a functional can be useful in time-series spectroscopy nonetheless.  We will present an example that shows small-p behavior by using a small positive fractional p-(quasi)norm, $L_{0.01}$.

The complex oscillatory behavior of fractional moments is very sensitive to the form of the distribution and can reveal properties of the distribution that are not otherwise readily apparent.  The oscillations can be analyzed by subsequently applying Fourier or wavelet transformations to identify changes in system behavior.

\subsection{Principal Components Analysis}

Principal components analysis (PCA) identifies linear subspaces that account for most of the variation in a data set.  This is tantamount to fitting a high-dimensional ellipsoid to the data.  This ellipsoid is also a contour of a multivariate Gaussian distribution fit to the data.  It is conventional to pre-process the data by subtracting the mean, such that the ellipsoid or Gaussian is centered at the origin.  Otherwise, the mean would generally be the first principal component.

PCA can be accomplished by either calculating the eigenvalues and eigenvectors of the data covariance matrix, or by directly calculating the singular value decomposition of the data matrix.  If the data matrix is $A$, then the data covariance matrix is $C=A^T A$.  $C$ can then be decomposed into a matrix of eigenvectors, $V$, and a diagonal matrix $\Sigma$ of eigenvalues:
\begin{equation}
C=V^T \Sigma V
\end{equation}

The n largest principal components are the eigenvectors associated with the n largest eigenvalues in $\Sigma$ multiplied by their weights, which are the square roots of these eigenvalues.  To obtain the low-rank PCA approximation of the covariance matrix, eigenvalues in $\Sigma$ less than the n largest are set to zero.

In addition to an approximation of the covariance matrix, a low-rank PCA approximation to the original matrix may be obtained via singular value decomposition.  The (zero-mean) data matrix may be decomposed into a diagonal matrix of singular values, $S$ and matrices of left and right singular vectors, $V$ and $U$, respectively:
\begin{equation}
A=V^T S U
\end{equation}

By setting the smallest eigenvalues in S to zero in the above expression, a low-rank approximation to the original matrix is obtained.  The principal components are the largest singular values multiplied by their corresponding left singular vectors.  The duals of these components may also be obtained by multiplying the largest singular values by the right singular vectors.  If the data matrix represents a time series, these duals indicate how the principal components change over time.

PCA is very useful in the context of spectroscopy and remote sensing\cite{boardman}, especially in complex or non-homogenous samples, because they identify spectral bands that vary together.  This can help to identify, for example, different minerals in a rock or different types of vegetation in a satellite image.

\subsection{Other approaches to parameterizing statistical distributions}

Another approach to a time-dependent characterization of the distribution would be to parameterize the distribution in terms of moments or cumulants and perform Bayesian inference using, e.g. Maximum Likelihood Estimation (MLE), or Maximum Posterior Probability (MAP).  One example of this would be to maximize the log-likelihood associated with a parameterized probability distribution such as:
\begin{equation}
f_X(x) =  Ce^{a_0+a_1 x+a_2 x^2+a_3 x^3+...+a_n x^n}
\end{equation}

In cases having a limited number of data (short sliding windows) MLE or MAP could potentially produce approximations of the probability distribution that are more accurate than directly calculating $E(X^n)$ applying their definitions to calculate moments and cumulants from the sample.

Many other techniques may be used to estimate moments or cumulants.  A Kalman filter, for example, could be used to estimate variances in spectral timeseries.

\section{Spectral Timeseries}

Since spectrometers typically collect a series of multiple scans and average them to produce spectra, calculating higher moments and cumulants of a series of scans may be accomplished through comparitively simple calculations, adding tremendous value to the spectroscopic information that they produce.  The nth moment may be readily calculated as the expectation of $X^n$ and the nth cumulant may be calculated from the first n moments.  Alternatively, these or other parameters of a probability density could be estimated via Bayesian inference or other methods.

Given a time-series of measurements of a particular spectral band, $S(t,\nu)$, we may separate these measurements into the band's mean (equilbrium) value, $\mu_t(t,\nu)$ and a zero-mean stochastic process $X(t,\nu)$, where $t=t_0, t_1, ... t_n$ is the set of time values over which the band has been sampled.  Each $X(t,\nu)$ is a random variable having a probability distribution with a mean of zero.  In this way,
\begin{equation}
S(t,\nu) = \mu(t,\nu) + X(t,\nu)
\end{equation}

Where $E(X(t,\nu)) = \int_{-\infty}^{\infty} X(t,\nu) dt = 0$.  The mean, $\mu(t,\nu)$ corresponds to the equilbrium 'graybody' spectrum, which may change with time.  The random variable $X$ describes departures from equilbrium, i.e. the distribution of non-equilbrium behavior of the spectrum.  Here, since the fluctuation from equilibrium, $X$, has zero mean, no distinction needs to be made between its raw moments and central moments.

Since the probability density associated with $X(t,\nu)$ is frequency- and time-dependent, its moments, cumulants, or other parameters are functions of both time and frequency to allow for statistical non-stationarity.  In some cases it may be useful to scale higher moments (or cumulants) by the standard deviation, producing standardized moments (or cumulants), equivalent to the moments (or cumulants) of the z-scores associated with the time series.

\subsection{Sliding-window approach}

If the timeseries is discretely sampled, as is often the case, and also nonstationary, it may not be possible to directly calculate the expectation values of powers of the underlying stochastic process.  In such a case, local estimates may be constructed.  One approach is to sample a local neighborhood of the timeseries, e.g. within a sliding window of width W, and to directly calculate moments or cumulants within this window.  
\begin{equation}
m_n(\nu,t,L) = \sum_{j=0}^W w_j X(t+j,\nu)^n
\end{equation}

Here $\sum_j w_j = 1$ and the unit of time is the timestep.  The $w_j$ could be uniformly distributed or could have a functional dependence, e.g. a binomial distribution.  The width of the sliding window is a tradeoff between temporal response or resolution, which favors smaller windows, and accurate estimation of moments from sampled values, which favors larger windows.  So long as the size of the window remains constant, the time complexity of such an approach is $O(n)$ and the method is suitable for real-time calculations.

Local estimation could also be accomplished via convolution with some compact kernel such as a Gaussian that moves in time but extends over over the entire signal.  The main disadvantage of such an approach is that the time complexity increases to $O(n^2)$, making the method less scalable.

From the first n time-dependent moments, the first n cumulants can be determined from the relations given earlier; these cumulants may have a more direct correspondence to extrinsic physical quantities due to their additive nature.  For example, all thermodynamic variables can be expressed in terms of cumulants of the free energy.  Moreover, in some cases, higher cumulants of energy can distinguish statistical mechanical systems that appear to be equivalent in terms of lower cumulants.

\subsection{Multi-scale approach}

In practice, a simple sliding window approach may work well for many applications, but it does introduce an additional free parameter to the problem - the length of the sliding window.  Without knowledge of the processes involved, the choice of an optimal window length may not be obvious.  Moreover, a sliding-window approach can effectively smooth the data, making it difficult to resolve features on timescales shorter than the window length.  A multi-scale approach solves both of these problems while maintaining linear time complexity.  For a sequence of length T, a level-j multi-scale parameterization of the time-varying moments is given by:
\begin{equation}
m_n(\nu,i,j) = \sum_{k=0}^{2^{-j}T-1} X(t+k+i2^{-j}T,\nu)^n
\end{equation}

Here, the index i runs from 0 to $2^l-1$.  A corresponding set of cumulants may be obtained in the usual way from a polynomial in the moments m(i,j).

\section{Example: Spectrum of CO2}

In order to validate the ability of higher time-dependent moments to extract emission lines, we consider the extraction of a carbon dioxide spectrum from ambient air without the use of any background spectra or curve-fitting.  A small notch in an averaged spectrum of ambient air can be seen in Figure 1 at 667cm$^{-1}$.  This corresponds to the CO$_2$ bending mode.

\begin{figure}
\centering
\includegraphics[width=9cm]{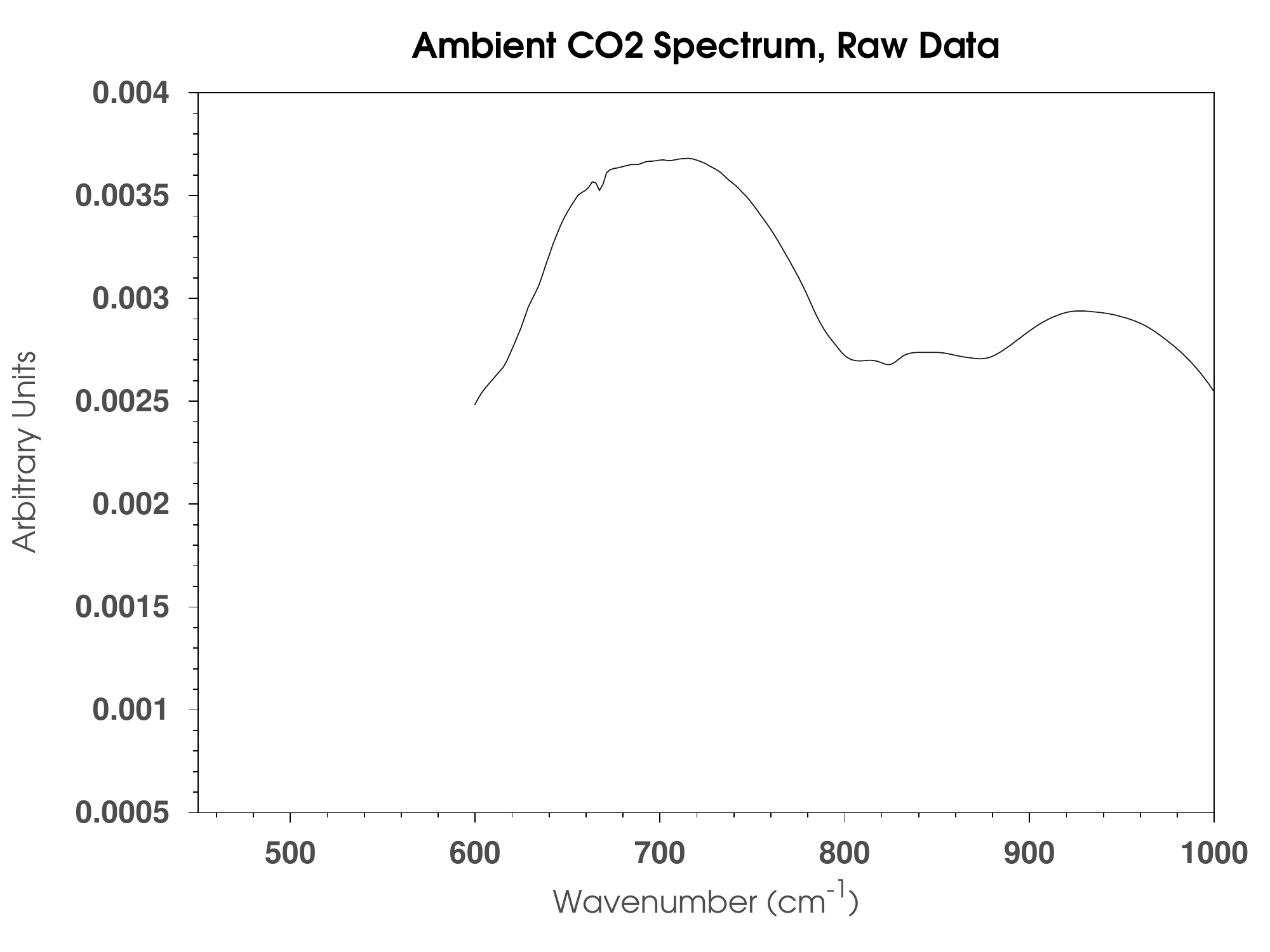}
\caption{Averaged emission spectrum of ambient air.  A faint CO$_2$ line can be seen at 667cm$^{-1}$.}
\end{figure}

However, the emission line associated has a significant amount of fine structure that is not apparent from this plot.  The following plot from the NIST-EPA webbook shows a high-quality reference IR absorbance spectrum for CO$_2$.  Distinct P- and R- branches are clearly visible, as are several peaks, in addition to the primary bending mode at 667cm$^{-1}$.

\begin{figure}[H]
\centering
\includegraphics[width=9cm]{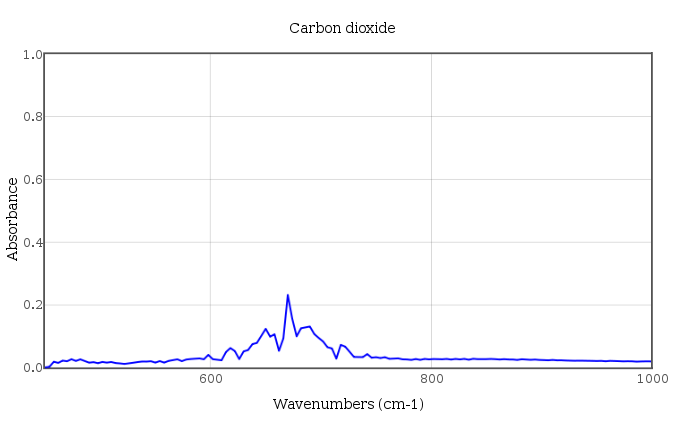}
\caption{Infrared CO$_2$ absorbance spectrum from the NIST Chemistry Webbook, NIST Standard Reference Database Number 69.  Reprinted courtesy of the National Institute of Standards and Technology, U.S. Department of Commerce.  Not copyrightable in the United States.}
\end{figure}

Due to Kirchhoff's law of thermal radiation, the emission spectrum is closely expected to match the absorbance spectrum.  

\subsection{Stochastic Time Series Spectroscopy}

We calculate the average of the fourth central moment (kurtosis) for each spectral band and take the fourth root such that the unit of the y-axis is linear in energy, like the original spectrum.  As expected, the result closely matches the reference spectrum.  In fact, its noise level is lower than that of the (circa 1964) reference spectrum, producing smoother P- and R- branches, and making the peaks associated with weak Q-branches more obvious.

\begin{figure}[H]
\centering
\includegraphics[width=9cm]{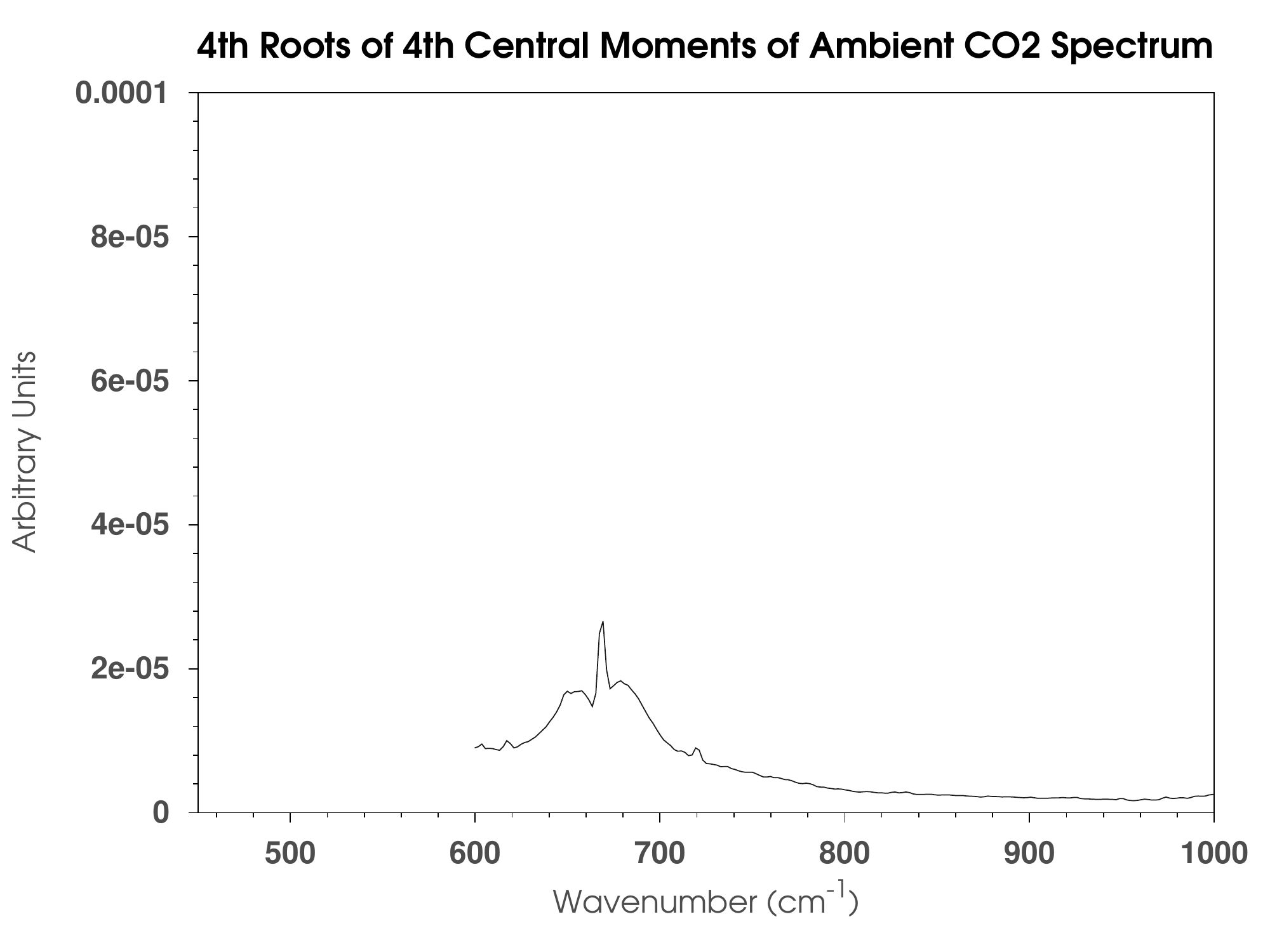}
\caption{Fourth root of Kurtosis for each band in the spectral time-series.  The result closely matches the reference spectrum.  Stochastic time-series spectroscopy can extract emission lines without any additional information.}
\end{figure}

Thus, stochastic time-series spectroscopy can extract emission lines without any additional processing.  Emission lines are intrinsically more random than the thermal equilibrium background, making them statistically distinct.

\subsection{PCA result}

For comparison, PCA was applied to the same data.  The first principal component shows the CO$_2$ line superimposed onto the thermal background with other features, like the water line.  The second principal component, however, compares favorably to both the reference sample and the STSS result.

\begin{figure}[H]
\centering
\includegraphics[width=9cm]{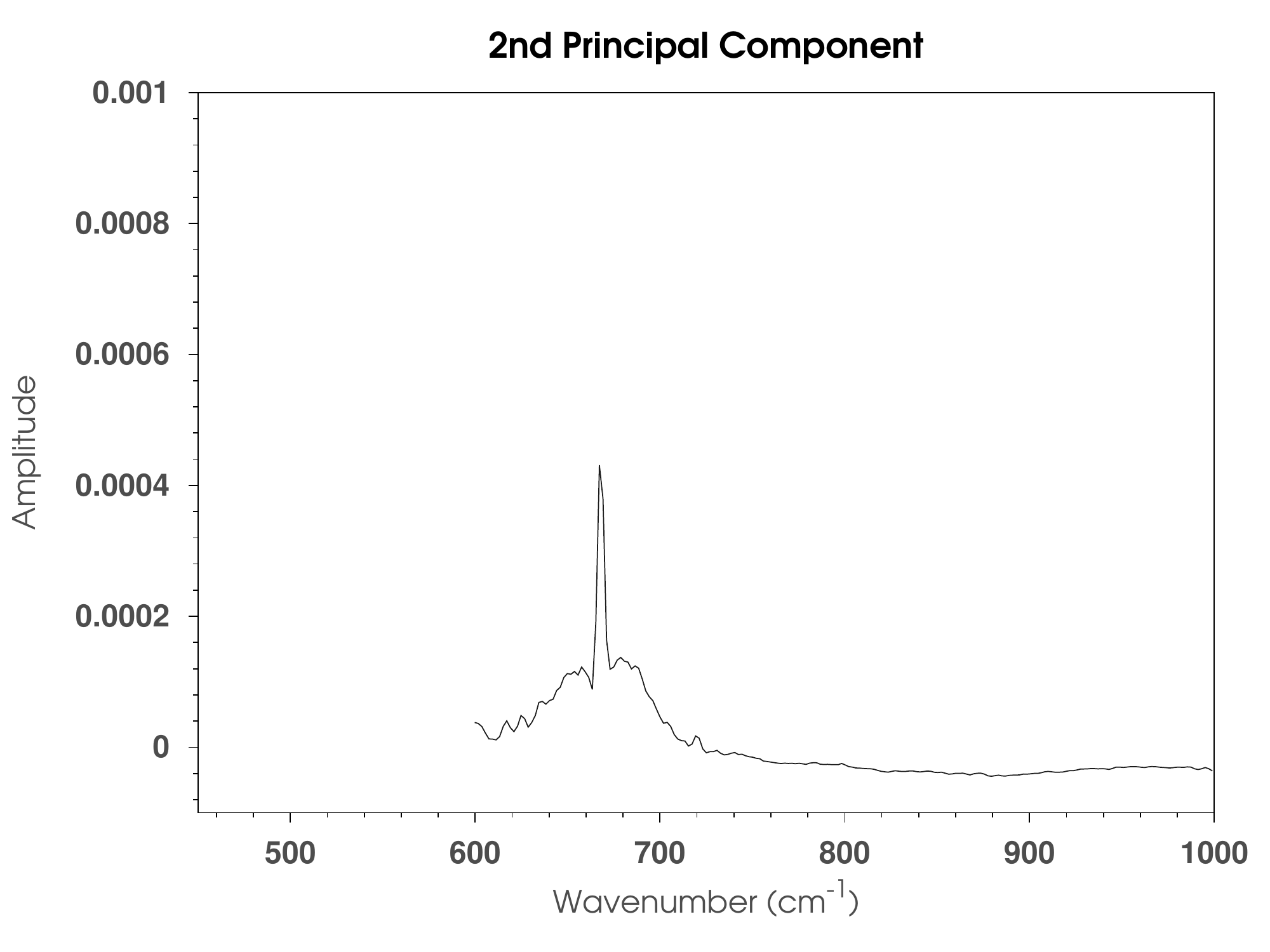}
\caption{The second principal component of the timeseries contains the CO$_2$ line.}
\end{figure}

Compared to PCA, STSS using the fourth root of kurtosis more closely matches the reference spectrum.  STSS yields a and lower-noise spectrum than either PCA or the NIST reference, more closely matching quantum mechanical predictions and high-resolution spectra.

\section{Example: observing stress in rocks}
 
We now consider a case study in which the concept of stochastic spectral time-series analysis is applied to regions where the wavelengths of interest are comparable to the wavelengths emitted by thermal radiation.  At temperatures that are typical to the Earth's surface, this band is the 'thermal infrared' (TIR) region from around 600cm-1 to 1200cm-1.  Stochastic Time-Series Spectroscopy was applied to identify stress-related TIR emission in rocks.  When rocks were subjected to stress, stochastic time series spectroscopy revealed that their spectrum changes in subtle but measureable ways.  The applied force can be seen below, in Figure 5, versus sample number.  About three seconds elapse between subsequent samples.  The displacement associated with deformation is shown in Figure 6.

\begin{figure}
\centering
\includegraphics[width=9cm]{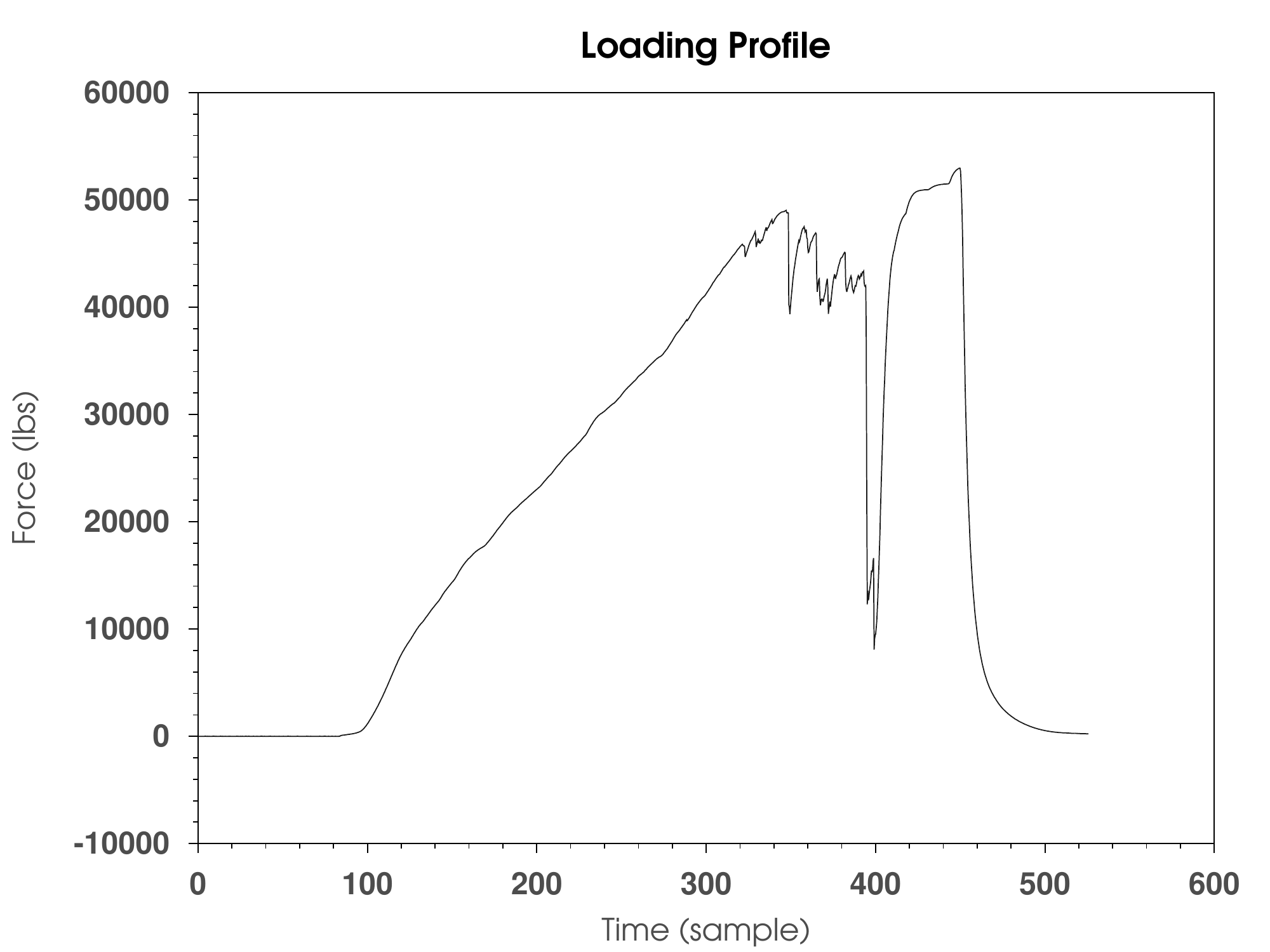}
\caption{Applied load as a function of time (sample number).}
\end{figure}

\begin{figure}
\centering
\includegraphics[width=9cm]{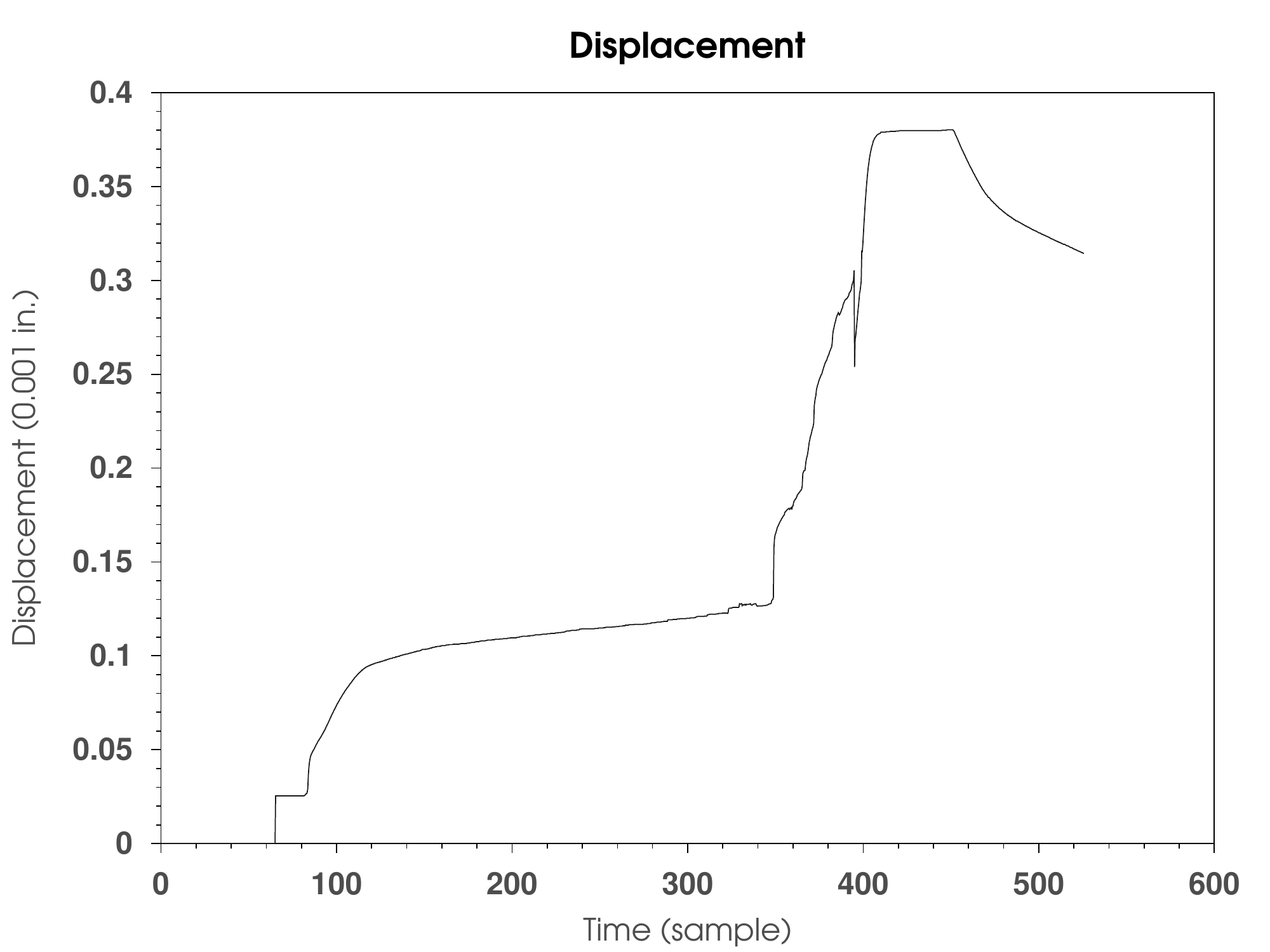}
\caption{Displacement of the hydraulic press as a function of time (sample number).}
\end{figure}

Fracture is evidenced by a rapid decrease of the applied load, or by rapid changes in displacement.  A series of smaller fracture events are apparent prior to total failure.  Spectra were collected at a location on the rock which was not stressed, so the relatively intense broad-spectrum radiation associated with fracture did not contribute to the collected spectra.  For this reason, changes in the rock's infrared spectrum were relatively subtle, requiring careful analysis.

The spectral data is characterized by a sequence of noisy oscillations in subsequent spectra.  It was determined that the amplitude and frequency of the oscillations changed once signficant pressure was applied to the rocks, but the common techniques of periodic analysis (e.g. Fourier transformations, autocorrelations, and wavelets) failed to resolve the noisy oscillations.  This was exacerbated by limited sampling frequency of the spectrometer, which was about 0.3$s^{-1}$.  The undersampling contributes to high levels of noise in the time series.  

Statistical methods, however, were very effective in analyzing the time series of spectral fluctuations.  By analyzing the statistical distribution of spectral fluctuations around their gray-body values, non-equilibrium spectral features previously buried in the thermal background were revealed at a high signal-to-noise ratio.

For comparison, principal components analysis was also applied to this data.  A purely PCA-based analysis was not able to resolve spectral features or resolve periods of stress in the time series.  However, by applying a variant of PCA in which PCA was applied within a sliding window, useful information about time-dependent behavior may be obtained.  Specifically, the weight of the first prinicipal component varies over time in a manner that agrees very well with the average of the time-dependent second moment.  To an extent, this is expected, since the component weights in PCA correspond to second moments of a multivariate Gaussian fit to the data.  Time averages of prinicipal components did not clearly resolve salient spectral features, although they did reveal some features, such as water lines, that were not visible in the non-time-varying principal components.

Stochastic time series spectroscopy was applied to a time-ordered sequence of spectra.  For each spectral band, a sliding window of fixed length was passed over the time series associciated with that band.  Within that band, the statistical distribution of spectral fluctuations was characterized the first four moments and cumulants.  Additionally, the p-norms $L_{50}$ and $L_{\infty}$, are presented, as well as the fractional p-(quasi)norm $L_{0.01}$.  This made it possible to spectroscopically determine when the rock was stressed or undergoing fracture, a characteristic that was not readily apparent from the raw spectral data.

Each of thse statistics provides a different lens through which to view spectroscopic data.  The differences observed between different moments, can reveal, for example, information about the statistics of emission events.  An example of this is a spectral line that exhibits a second moment, but no obvious higher moments.  This indicates that the emission is an approximately Gaussian process.  Conversely, an emission line that shows higher moments has a particular type of non-Gaussian time dependence.  In some cases (fractional p-quasinorms, for example) the interpretation of these statistics may not be readily apparent, but they may still be useful for detecting changes in the behavior of a system. 

\begin{figure}[H]
\centering
\includegraphics[width=9cm]{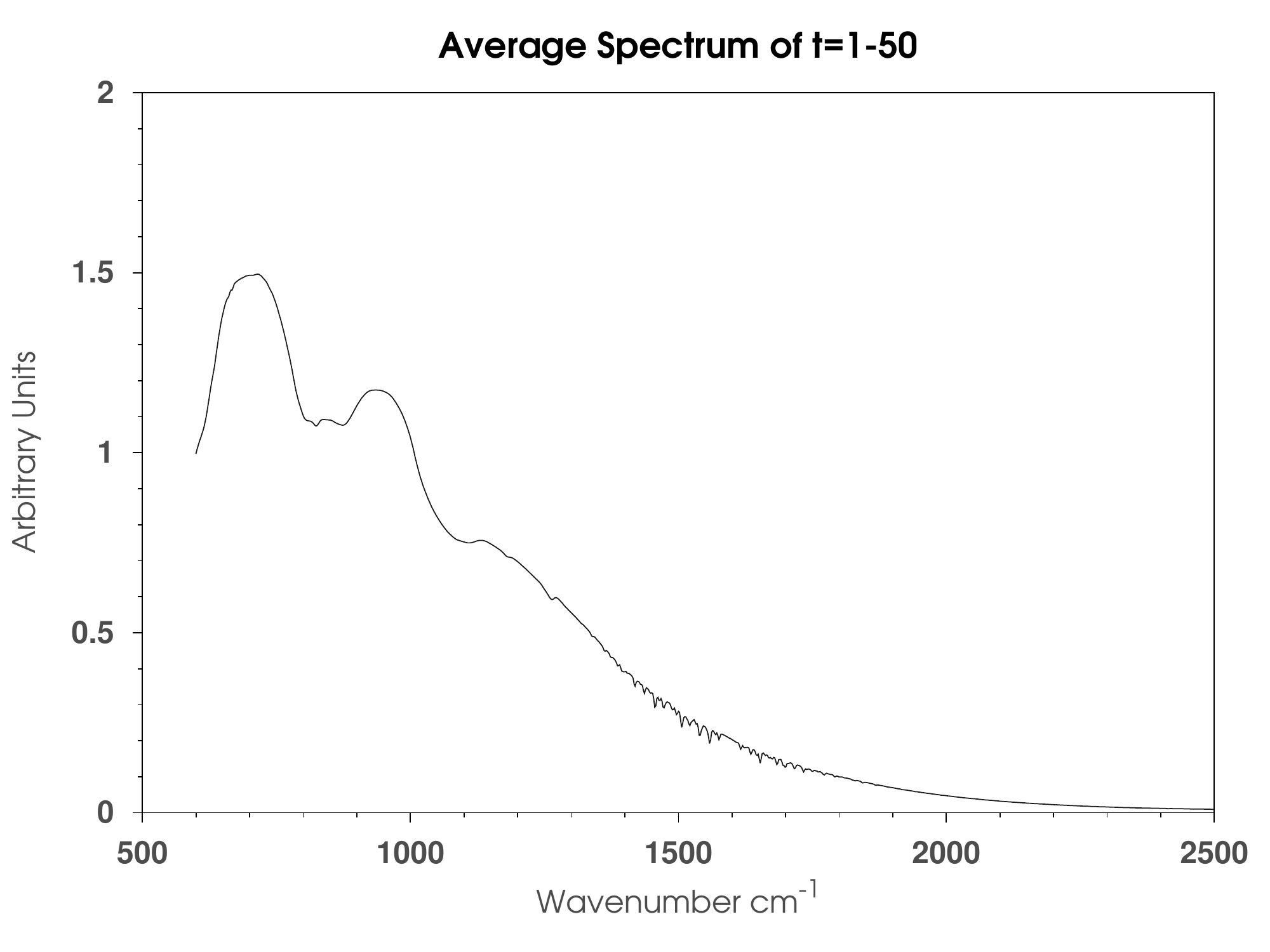}
\includegraphics[width=9cm]{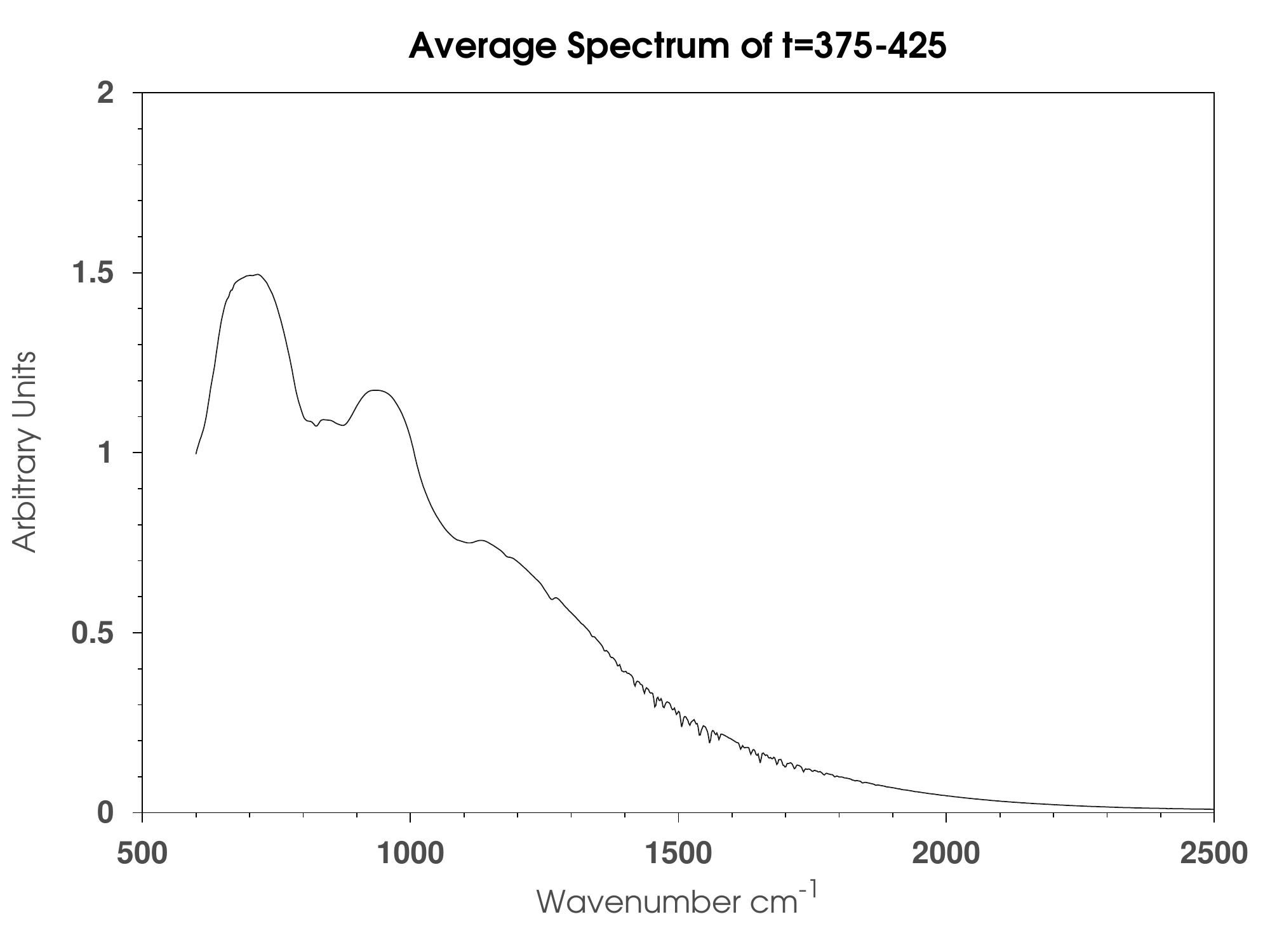}
\caption{Infrared emission from a sample of Red Granite (and ambient atmospheric gases) from 600 cm$^{-1}$ to 2500 cm$^{-1}$ before (above, average spectrum from t=1-50) and after (below, average spectrum from t=375-425) the application of catastrophic stress that pulverized the rock.  The two spectra are indistinguishable.}
\end{figure}

\subsection{The first moment and cumulant: the mean}

Simply measuring the TIR emission with a spectrometer did not reveal any obvious changes in the graybody spectrum seen by the spectrometer (in the presence of atmospheric gases), as evidenced by the following spectra, taken before and after the application of stress to a sample of red granite, shown in Figure 7.

\begin{figure}[H]
\centering
\includegraphics[width=9cm]{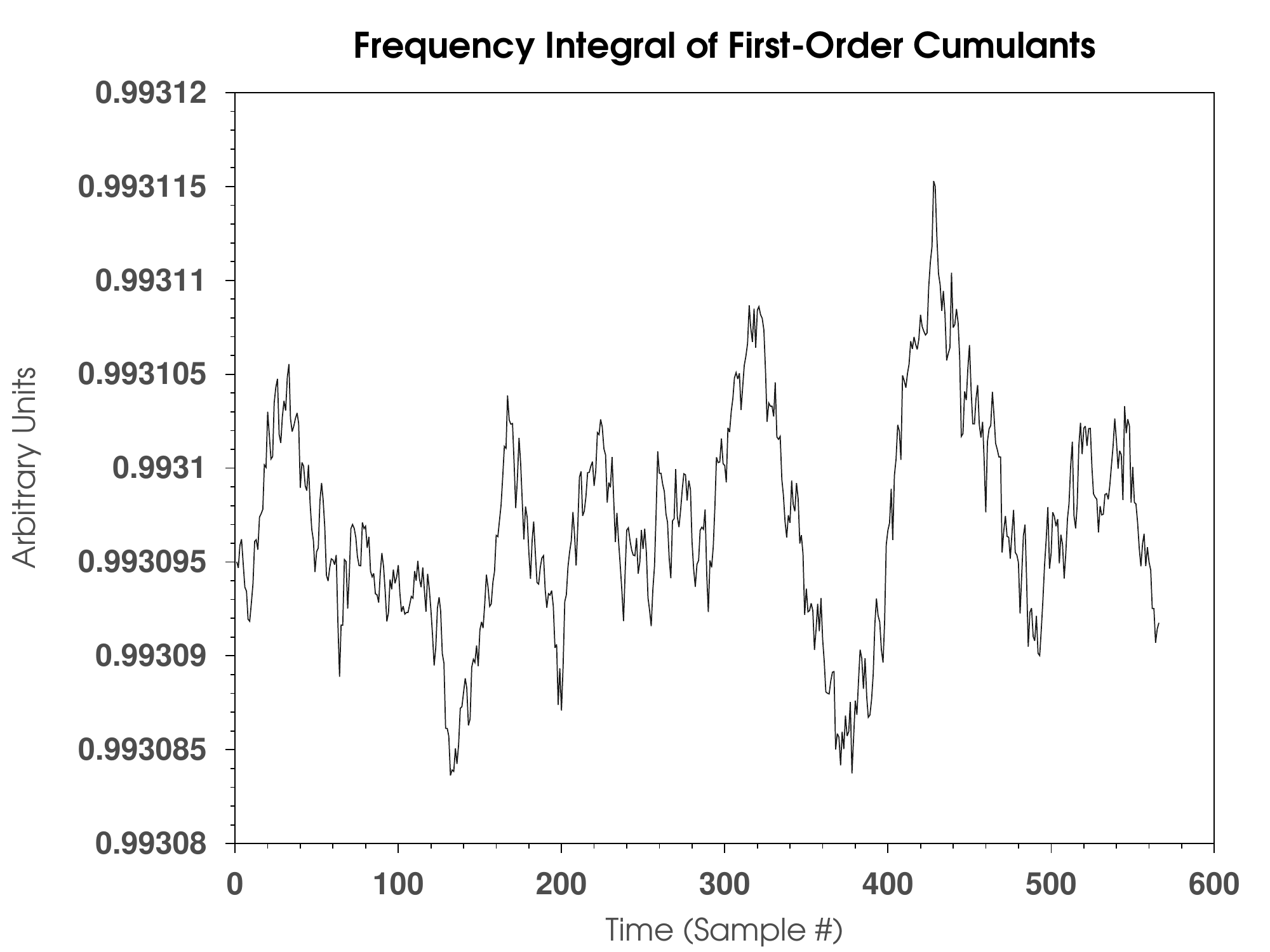}
\caption{The sum over frequency bands from 600 cm$^{-1}$ to 2500 cm$^{-1}$ of a moving average taken between a spectrum at time $t$ and time $t+30$.  Some noisy but weakly periodic behavior (correlated or colored noise) seems to be present.  A change of state is difficult to discern.}
\end{figure}

A non-equilibrium effect was not apparent from the raw spectra, which reveal little about the state of the rock.  Time-series of the spectra were available, so a moving average (an estimate of the time-depenent first raw moment) was calculated.  Since the spectra are not changing significantly, the moving average of the total intensity integrated over the 600 cm$^{-1}$ to 2500 cm$^{-1}$ band, shown in Figure 8, is relatively noisy. The time integral of the moving average simply recovers the spectrum, which is visually indistinguishable from those shown in Figure 7.

Examining difference spectra taken between a spectrum at time t and this moving average, the system seemed to exhibit some noisy periodic behavior, but these oscillations were not readily quantified by Fourier analysis, generalized harmonic analysis, or wavelet transformations.  This was due to a relatively high noise level and undersampling of the data.  These techniques revealed little information about the stress being applied to the rock.

\subsection{The second moment and cumulant: the variance}

The changing amplitude of noisy spectral fluctuations are captured in a robust manner by calculating the standard deviation of each spectral band within the sliding window, which is the square root of the second moment or cumulant.  This reveals how much each spectral band is fluctuating about its equilibrium value.  

The (time-dependent) second moment was calculating within a sliding window of width 30.  Integrating the time-dependent variance over the frequency bands from 600 cm$^{-1}$ to 1300 cm$^{-1}$, we may quantify the total TIR-band deviation from equilibrium as a function of time.  Plotting the standard deviation, the square root of variance, the spectral fluctuations due to hightened stress levels in the rock become quite evident, as shown in figure 9.

\begin{figure}
\centering
\includegraphics[width=9cm]{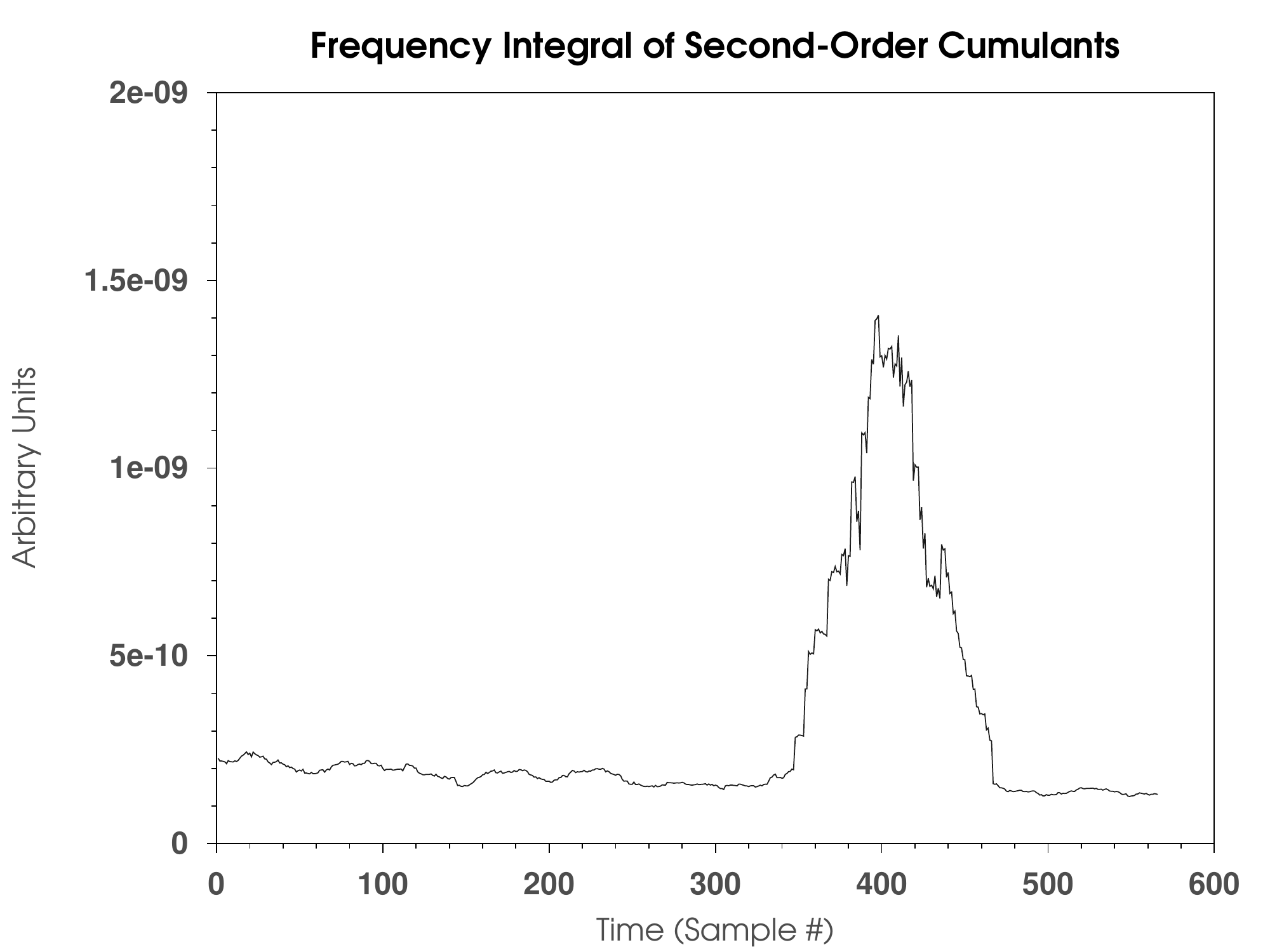}
\caption{The sum of the standard deviations of spectral bands, within a sliding window from sample $t$ to $t+30$.  Spectral fluctuations due to non-equilibrium behavior between samples $t=340$ and $t=460$ are clearly revealed by the analysis of the second central moment/cumulant.  The rock was pulverized due to stress near the maximum in the above plot, around $t=400$, as recorded by a pressure transducer.  This was not readily apparent from the raw data.}
\end{figure}

Integrating the time-dependent standard deviation of the process over time for each spectral band, we obtain the following plot.  The shape of this curve shows the average of the second moment over the course of the experiment.

\begin{figure}
\centering
\includegraphics[width=9cm]{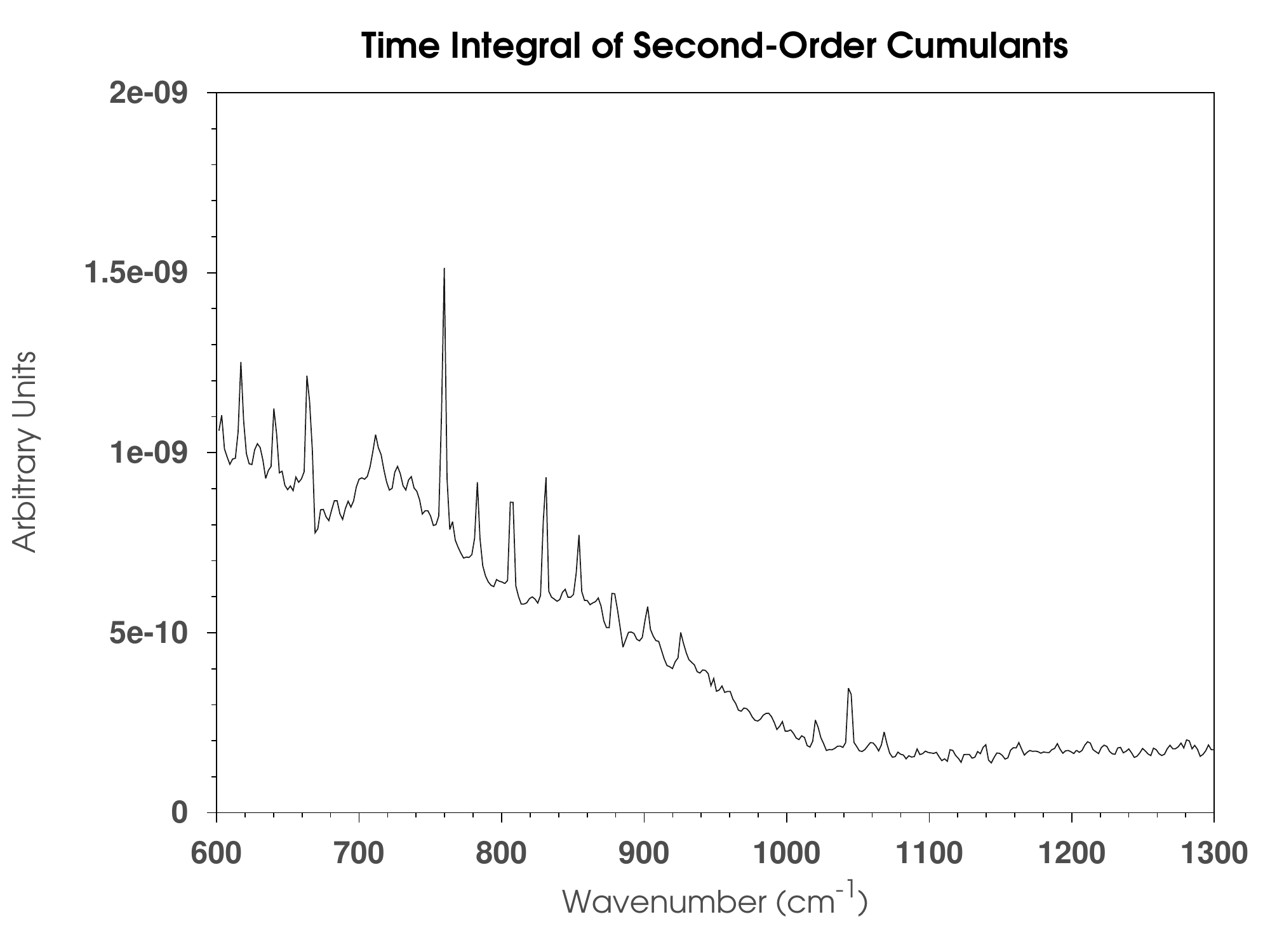}
\caption{The sum over all time of the standard deviation of each spectral band, where the standard deviation is evaluated within a sliding window from sample $t$ to $t+30$.  Note the presence of sharp emission/absorption lines.}
\end{figure}

Note that these should not be interpreted in the same manner as a difference spectrum, e.g. the standard deviation is always positive.  Nonetheless, it clearly shows the location of several very sharp spectral lines that were not apparent in the raw data, as well as a broad increase in spectral variance from 600 cm$^{-1}$ to 1000 cm$^{-1}$, indicating thermodynamic fluctuations in these bands.  

Larger values in these plots indicate that, at certain times, particular spectral bands have an increased tendency to deviate from gray-body equilibrium values.  In other words, non-equilibrium behavior is revealed.  When analyzed in this way, intrinsically non-equilibrium features such as emission and absorption lines are enhanced, while 'equilbrium' features such as relatively slow changes in temperature are suppressed.

\subsection{Comparison to PCA}

Before proceeding to higher moments, cumulants, and norms, the performance of PCA on these data should be considered.  Like STSS using the second moment/cumulant (the variance) a zero-mean PCA approximation describes a Gaussian distribution having only a second.  

The difference between PCA and second-order STSS lies in the fact that PCA describes a multivariate Gaussian distribution that typically has a dimensionality less than the number of spectral bands, while STSS is the cartesian product of a 1-D gaussian for each spectral band, or, equivalently, a multidimensional Gaussian whose dimensionality is equal to the number of spectral bands, and that is also orthogonal to the spectral bands.  

For these reasons, the two methods will produce similar results for certain problems.  For the example of the stressed rocks, however, the principal components reveal no obviously discernable spectral features and reveal nothing about temporal behavior.  The principal component, shown in Figure 11, is very noisy.

\begin{figure}
\centering
\includegraphics[width=9cm]{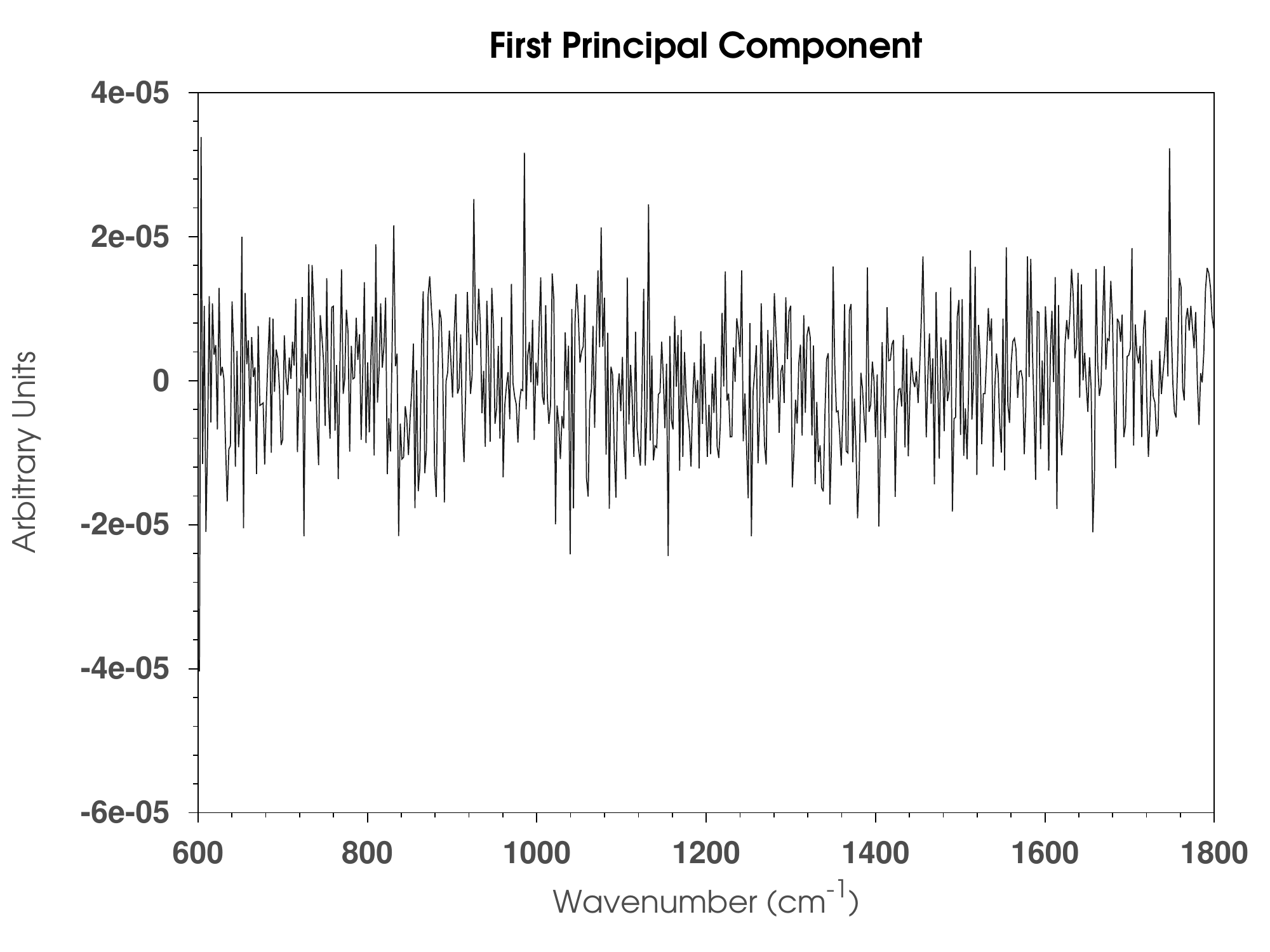}
\caption{The first principal component of a spectral time-series taken from the stressed rock.  It has a high noise level.  Subsequent principal components behaves similarly to the first.}
\end{figure}

If, on the other hand, PCA analysis is done in a sliding window to obtain a time-series of PCA approximations, the situation improves somewhat.  The time average of the first principal component calculated within a 50-sample sliding window can be seen below.  A sharp cutoff around 700cm$^{-1}$ can be seen.  This cutoff was present in all averaged PCA spectra in this study, including controls, and is presumably an artifact due to rapidly decreasing sensitivity of the Hg-Cd-Te detector below this level.

\begin{figure}[H]
\centering
\includegraphics[width=9cm]{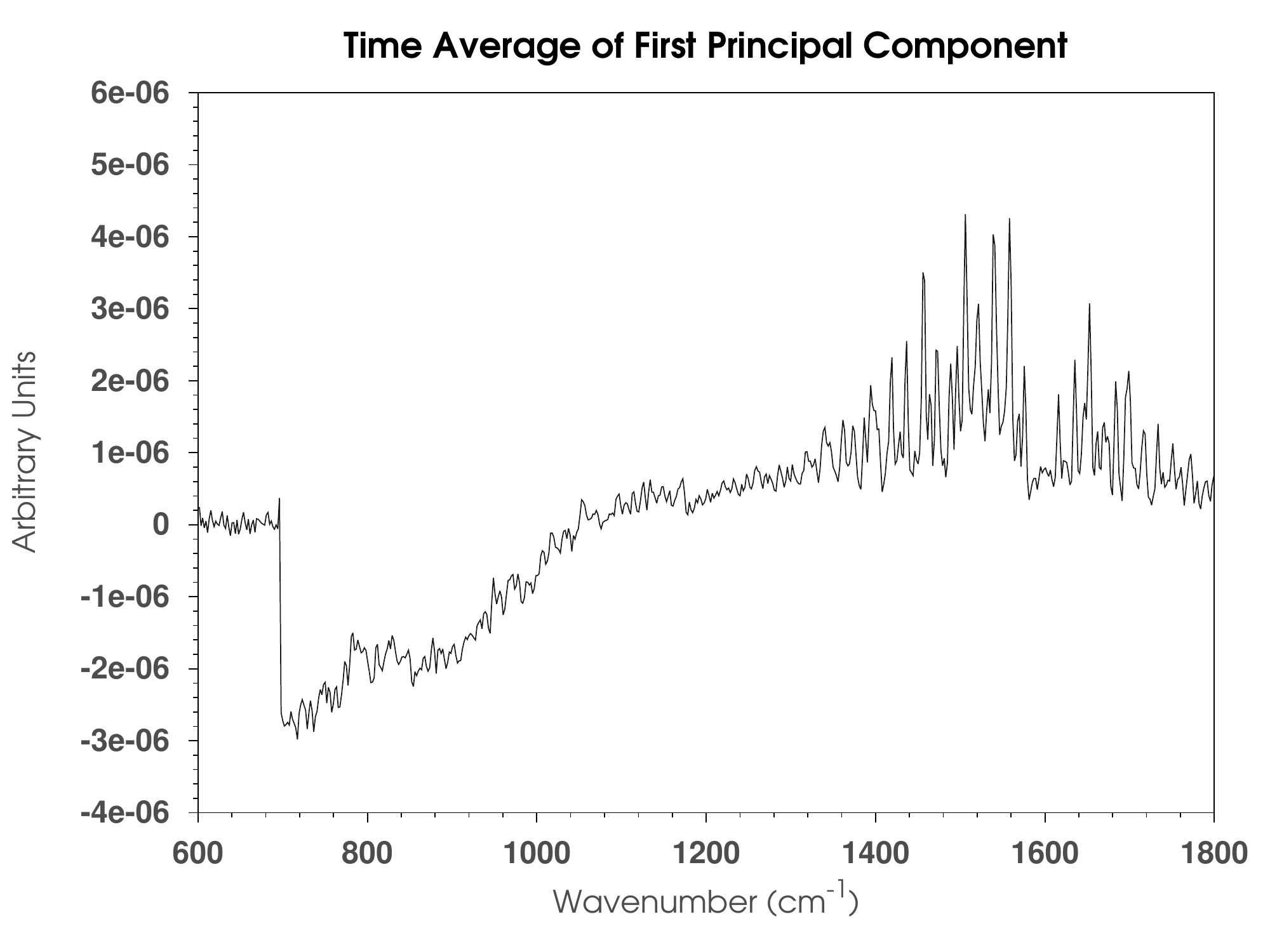}
\caption{The time average of the first principal component, where PCA was calculated in a 50-sample sliding window.  The noise level is reduced, and, ignoring that it is upside-down relative to the strictly-positive second-order STSS, it shows a similar decrease in amplitude from 700cm$^{-1}$ to 1000cm$^{-1}$.  It does not show sharp spectral lines, but does show a sharp cutoff just below 700cm$^{-1}$.  This cutoff is present in all principal components, and is presumably due to rapidly decreasing sensitivity of the Hg-Cd-Te detector below this level.  }
\end{figure}

If one considers the time series of component weights from sliding-window PCA, the result is very similar to variance-based second-order STSS.  This is to be expected since the major axis of the PCA ellipsoid typically accounts for a large fraction of the total variance.  Its computation time was vastly greater, however.  The computational time of PCA in a window of constant length is roughly $O(n^3)$, while STSS is roughly $O(n)$, so if the window size scales with the number of samples, PCA takes time rougly proportional to the cube of that of STSS.

\begin{figure}[H]
\centering
\includegraphics[width=9cm]{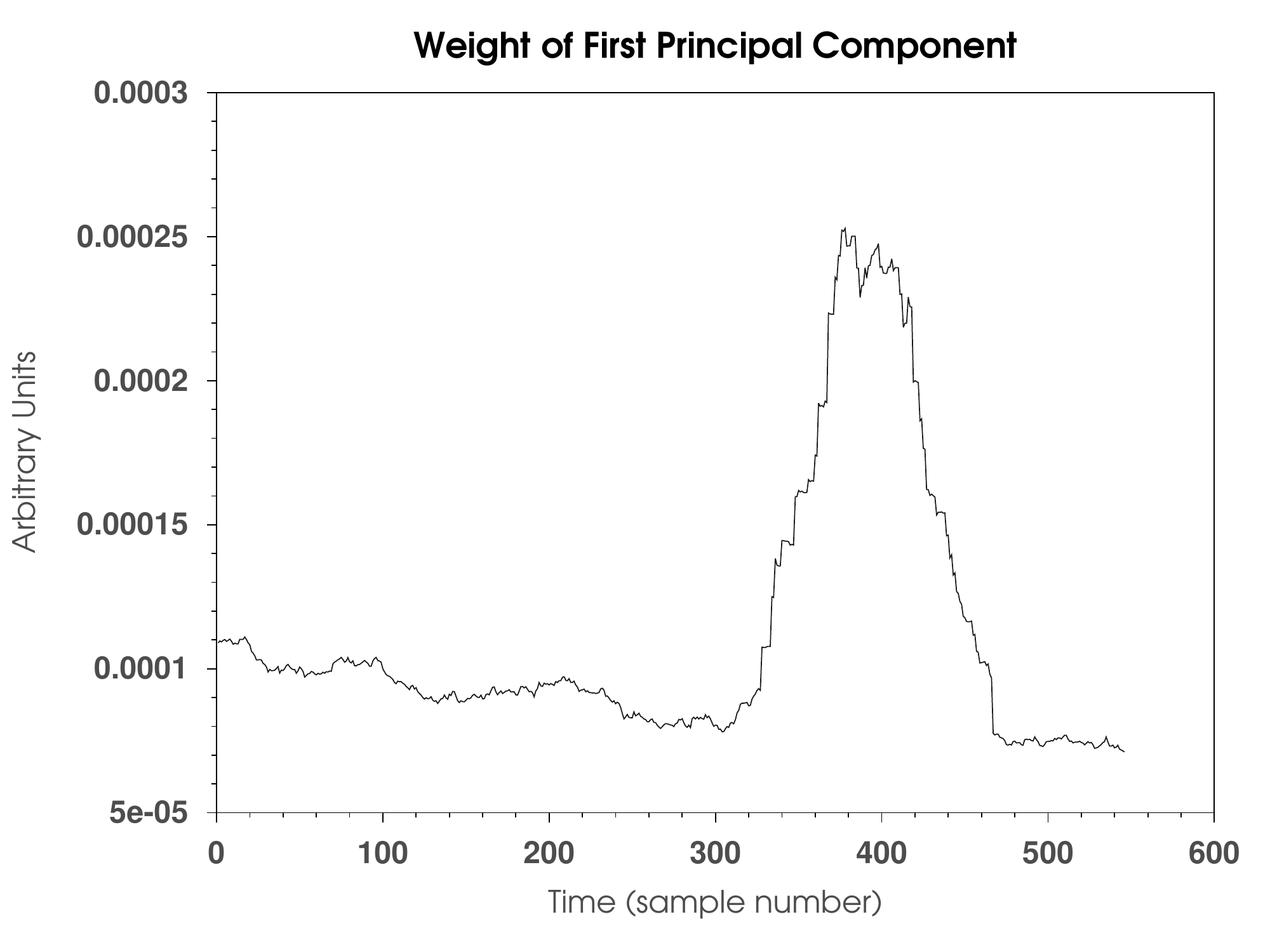}
\caption{The weight of the first principal component, where PCA was calculated in a 50-sample sliding window.  This closely follows the frequency averaged second moment/cumulant, as expected, since the major axis of the PCA ellipsoid accounts for the largest fraction of the variance.}
\end{figure}

\subsection{The third moment and cumulant: the skewness}

Higher moments and cumulants can potentially reveal more information, providing better time resolution or signal-to noise ratio, even distinguishing phenomena that appear similar given lower-order statistics.  The third moment about the mean is the skewness, showing whether a distribution tends toward high or low values.  Unlike the standard deviation, which is always positive, the sign of skewness (again calculated in a sliding window of width 30) can reveal direction in the system, as shown in figure 14.  

\begin{figure}
\centering
\includegraphics[width=9cm]{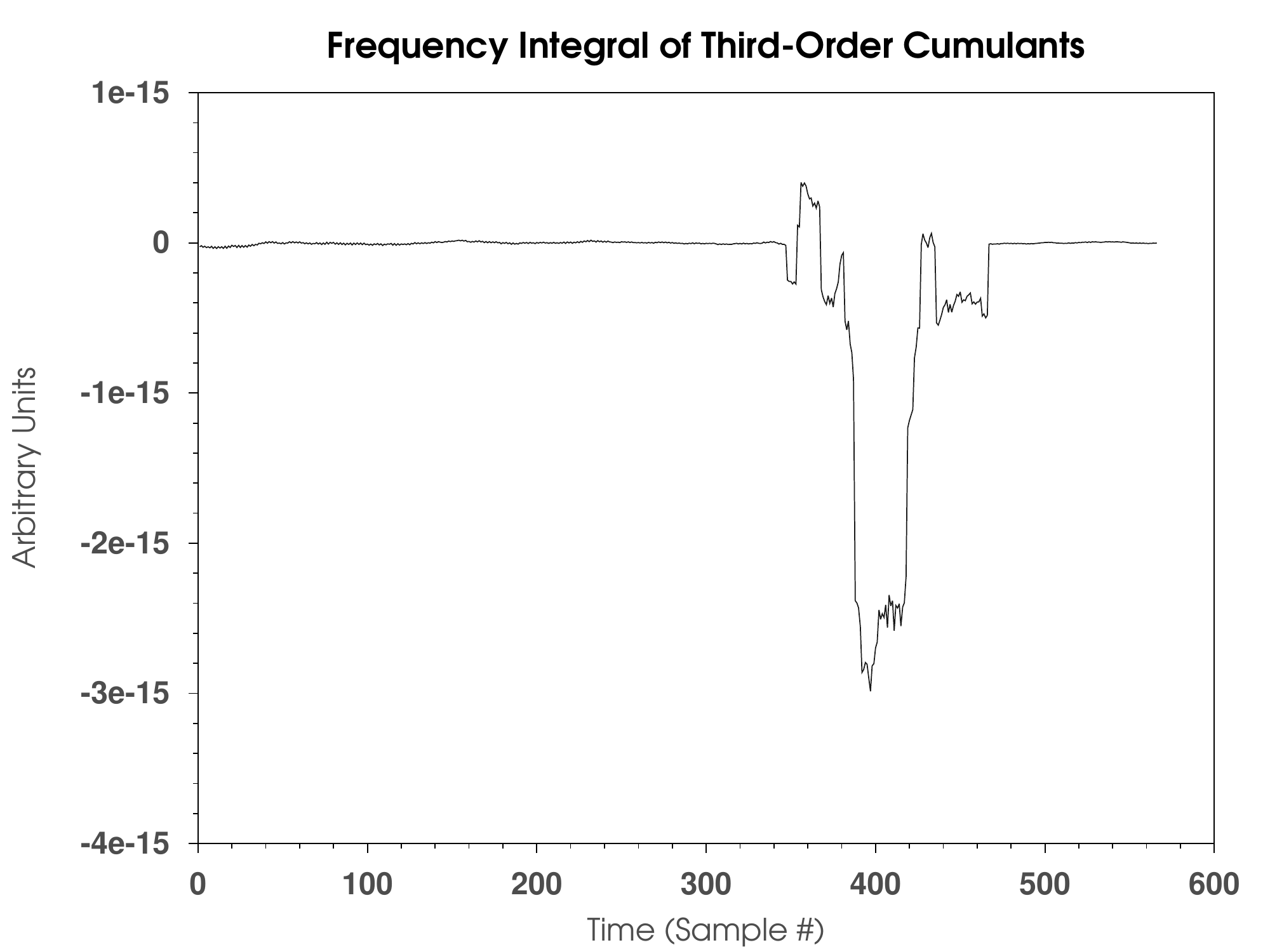}
\caption{The sum of the skewness of spectral bands, within a sliding window from sample $t$ to $t+30$.  Analysis via the third central moment/cumulant not only clearly identifies non-equilibrium states with an excellent signal-to-noise ratio, but it seems to differentiate between IR emission during the onset of fracturing (anomalous positive net skewness) and failure (anomalous negative net skewness) non-equilibrium behavior.}
\end{figure}

In the case of the rock, a non-equilbrium state of extremely positive skewness can be seen prior to the initial onset of fracturing, well before total failure occurs, whereas a state of extremely negative skewness can be seen after the initial onset of fracturing.  Analysis of higher moments/cumulants of the spectral time-series, such as the skewness, could potentially change sign in response to a slight changes in state in a material.  Integrating the skewness over time rather than frequency reveals the mean skewness for each spectral band, shown in Figure 15.  

\begin{figure}[H]
\centering
\includegraphics[width=9cm]{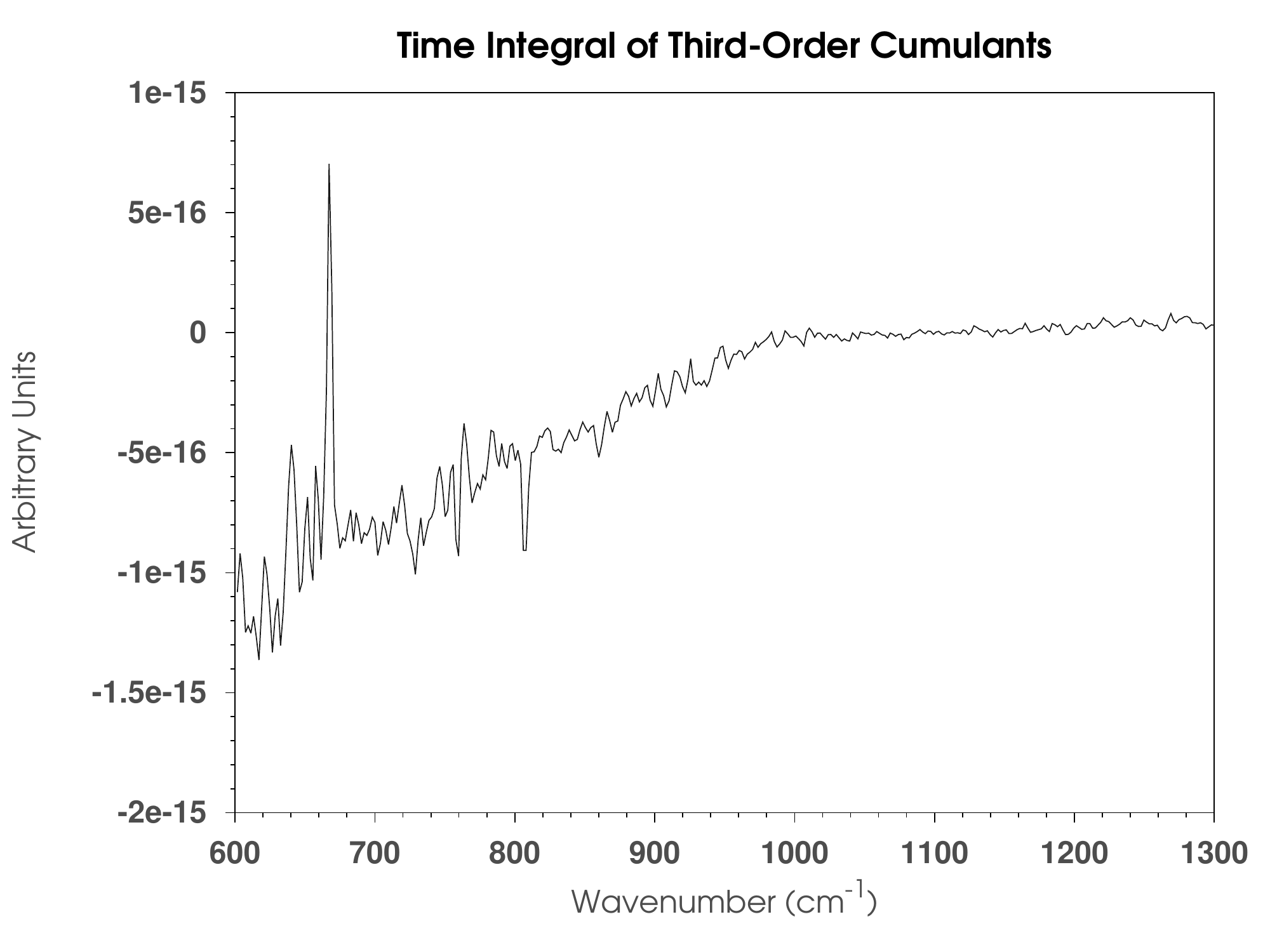}
\caption{The sum over all time of the skewness of each spectral band, where the skewness is evaluated within a sliding window from sample $t$ to $t+30$.  Note the presence of sharp emission/absorption lines, both upward and downward.}
\end{figure}

Figure 15 shows strongly negative skewness from 600 cm$^{-1}$ to 1000 cm$^{-1}$, indicating statistical fluctuations in this region, as could also be seen via the second moment/cumulant.  Compared to the second moment, however, the third moment/cumulant offers a different view of the spectral lines - in addition to a sharply elevated CO$_2$ line and somewhat elevated water lines, additional sharp lines are present which point downward.

\subsection{The fourth cumulant: excess kurtosis}

The fourth cumulant is the excess kurtosis, whereas the fourth central moment is the kurtosis.  Unfortunately, these terms are sometimes used interchangably.  The fourth cumulant differs from the fourth central moment by a factor of three times the square of the second cumulant.  

Considering the fourth cumulant associated with the stressed granite sample, the excess kurtosis of the spectral time series, a clear difference can be seen before and after catastrophic failure around t=400.  Here, stress in the rock has the effect of generating positive excess kurtosis.  

From the excess kurtosis, evidence of a secondary fracture event can be seen after the initial failure, after which the rock was re-loaded.  Additionally, smaller maxima can be seen during the rising action of the excess kurtosis, possibly indicating that stress was being released by microfracturing periodically during the loading process.  This is quite plausible, as the rock was noted to make crackling cracking sounds prior to failure, and in some cases, these were registered by the pressure transducer.
\begin{figure}
\centering
\includegraphics[width=9cm]{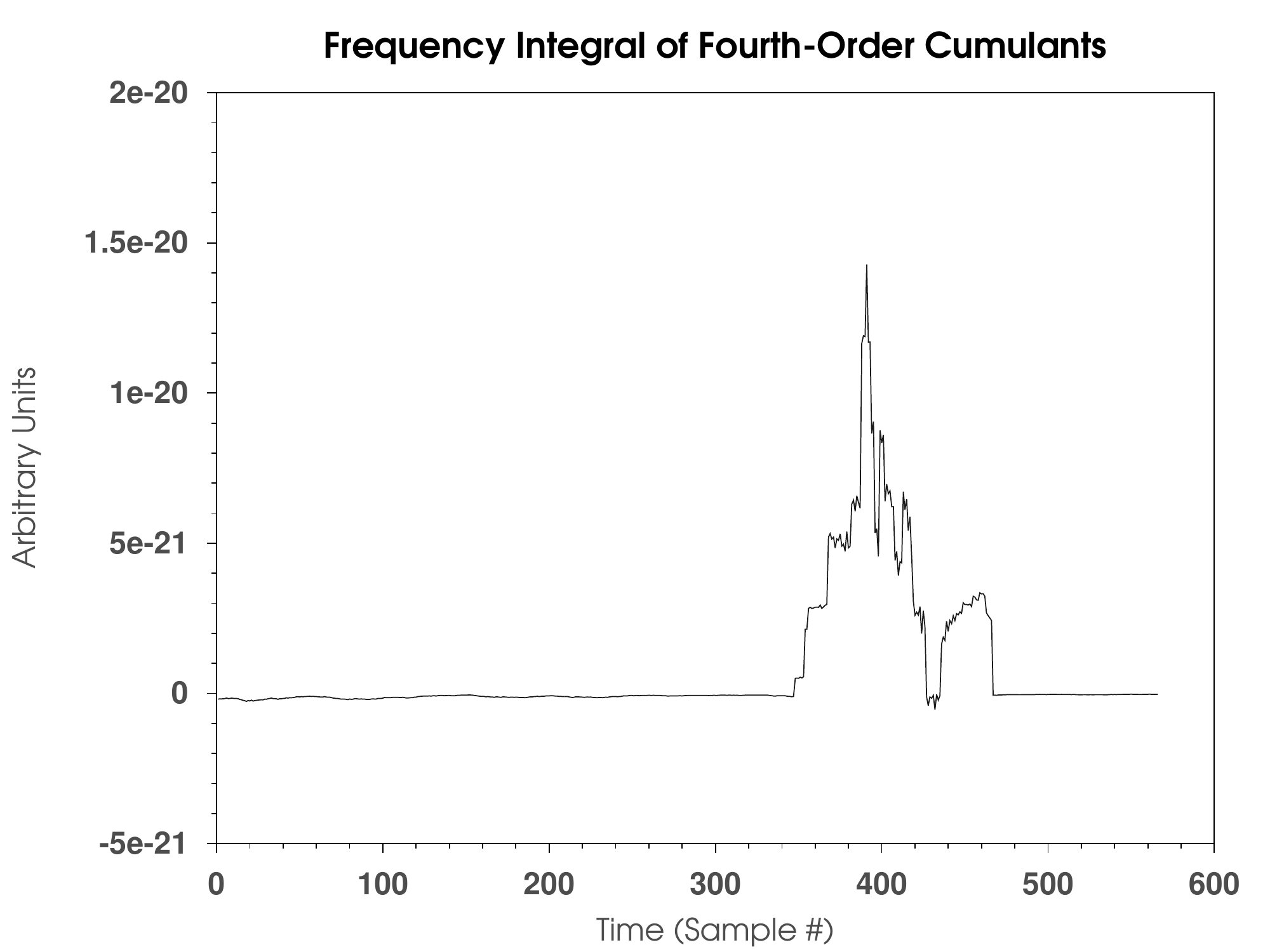}
\caption{The sum of the excess kurtoses of spectral bands, within a sliding window from sample $t$ to $t+30$.  Analysis via the fourth cumulant clearly shows the build-up to failure and the release of stress after failure.}
\end{figure}

\begin{figure}
\centering
\includegraphics[width=9cm]{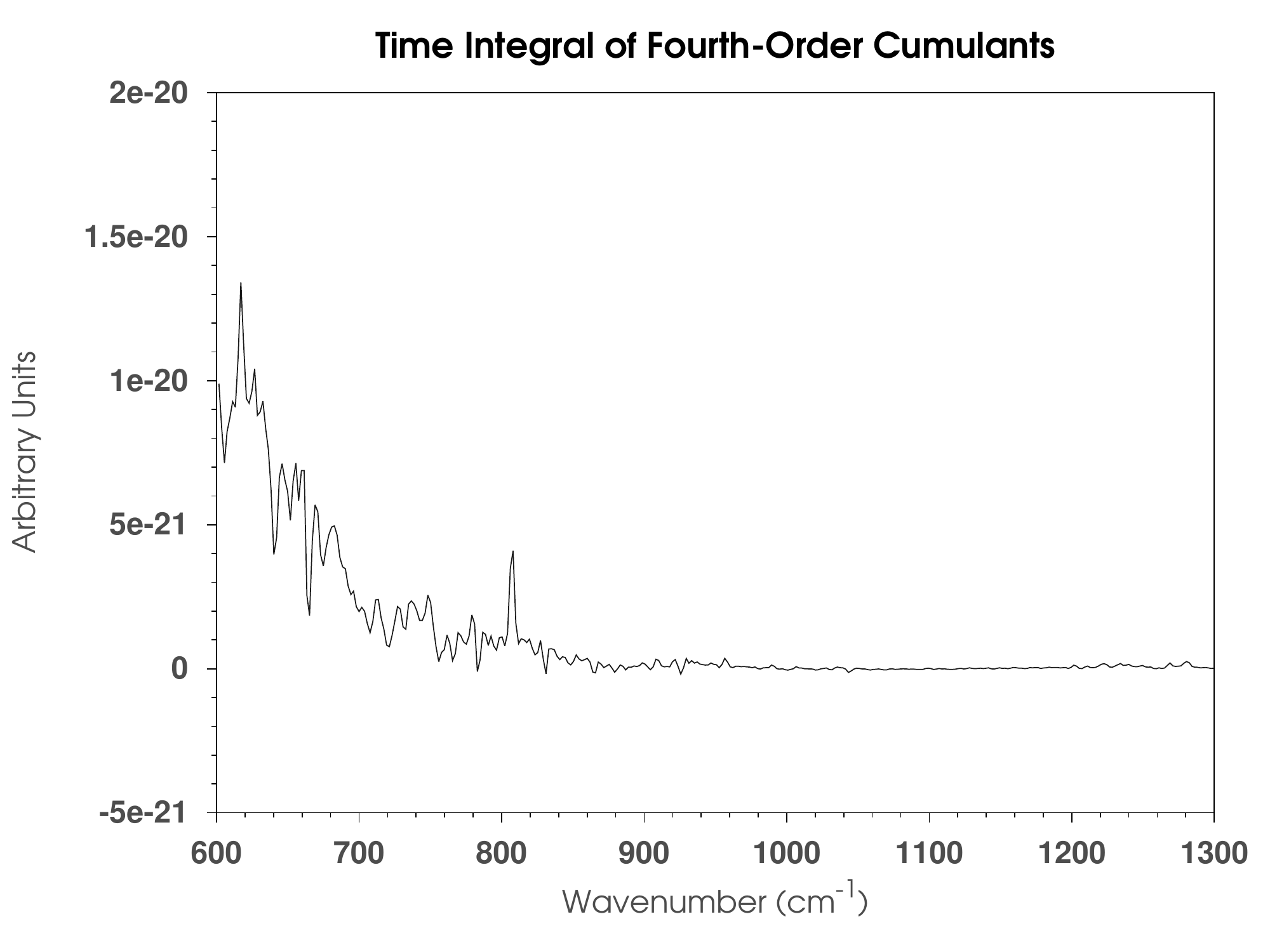}
\caption{The sum over all time of the excess kurtosis of each spectral band, where the excess kurtosis is evaluated within a sliding window from sample $t$ to $t+30$.  Note the presence of sharp emission/absorption lines, both upward and downward.}
\end{figure}

Integrating each band over time, we arrive at Figure 17.  Again, very sharp peaks are present, rising far above the low background noise level.  

\subsection{The fourth moment: kurtosis}

The fourth central moment is the kurtosis, showing how flat or peaked a distribution tends to be.  Both of these statistics are useful identifying whether bands have a simple or complex time dependence, hinting at the randomness associated with absorption and emission lines.

\begin{figure}
\centering
\includegraphics[width=9cm]{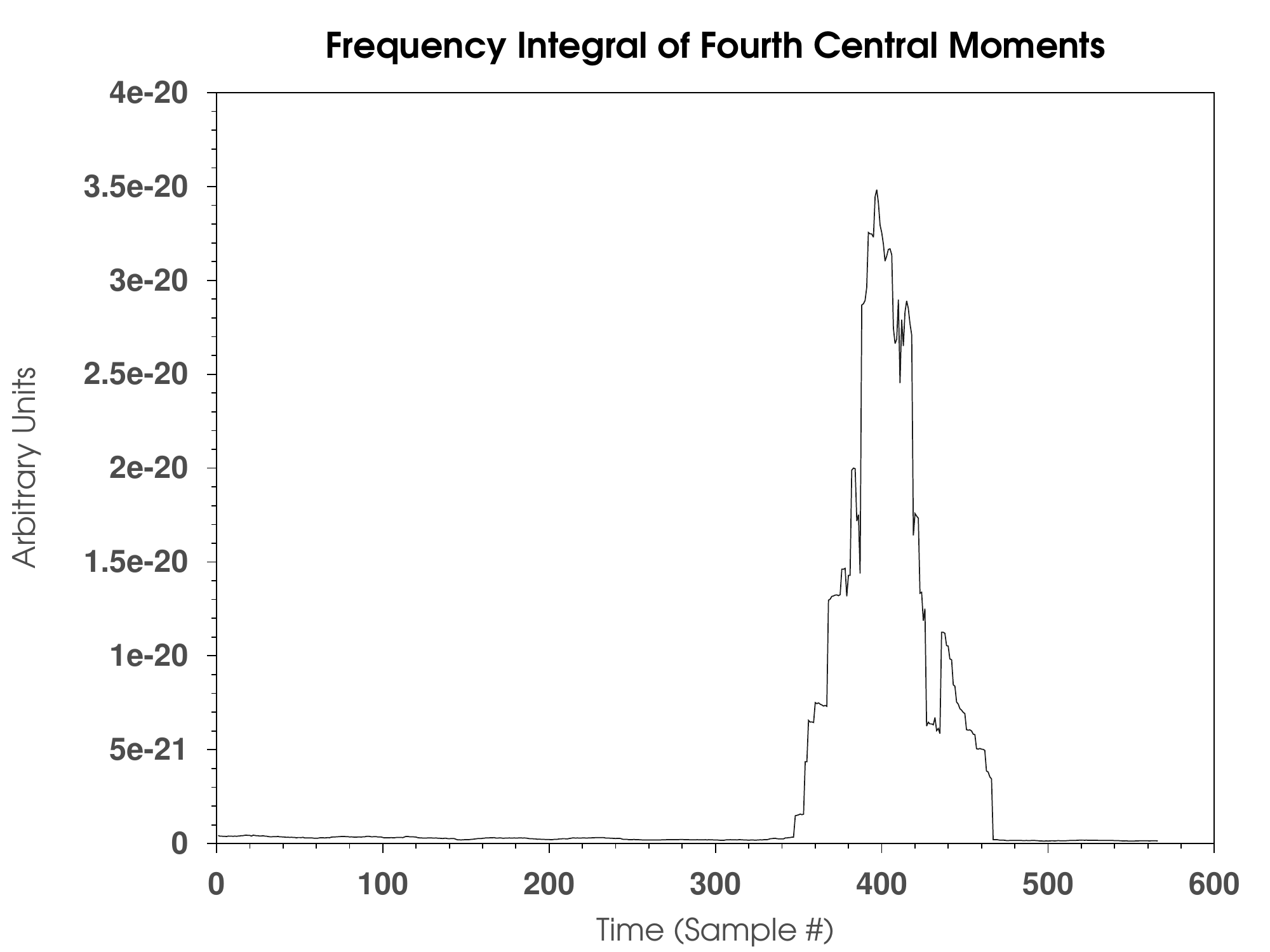}
\caption{The average of the kurtoses of spectral bands, within a sliding window from sample $t$ to $t+30$.  Like the fourth cumulant, the fourth moment clearly shows the build-up to failure and the release of stress after failure, albeit with a different profile.}
\end{figure}

\begin{figure}
\centering
\includegraphics[width=9cm]{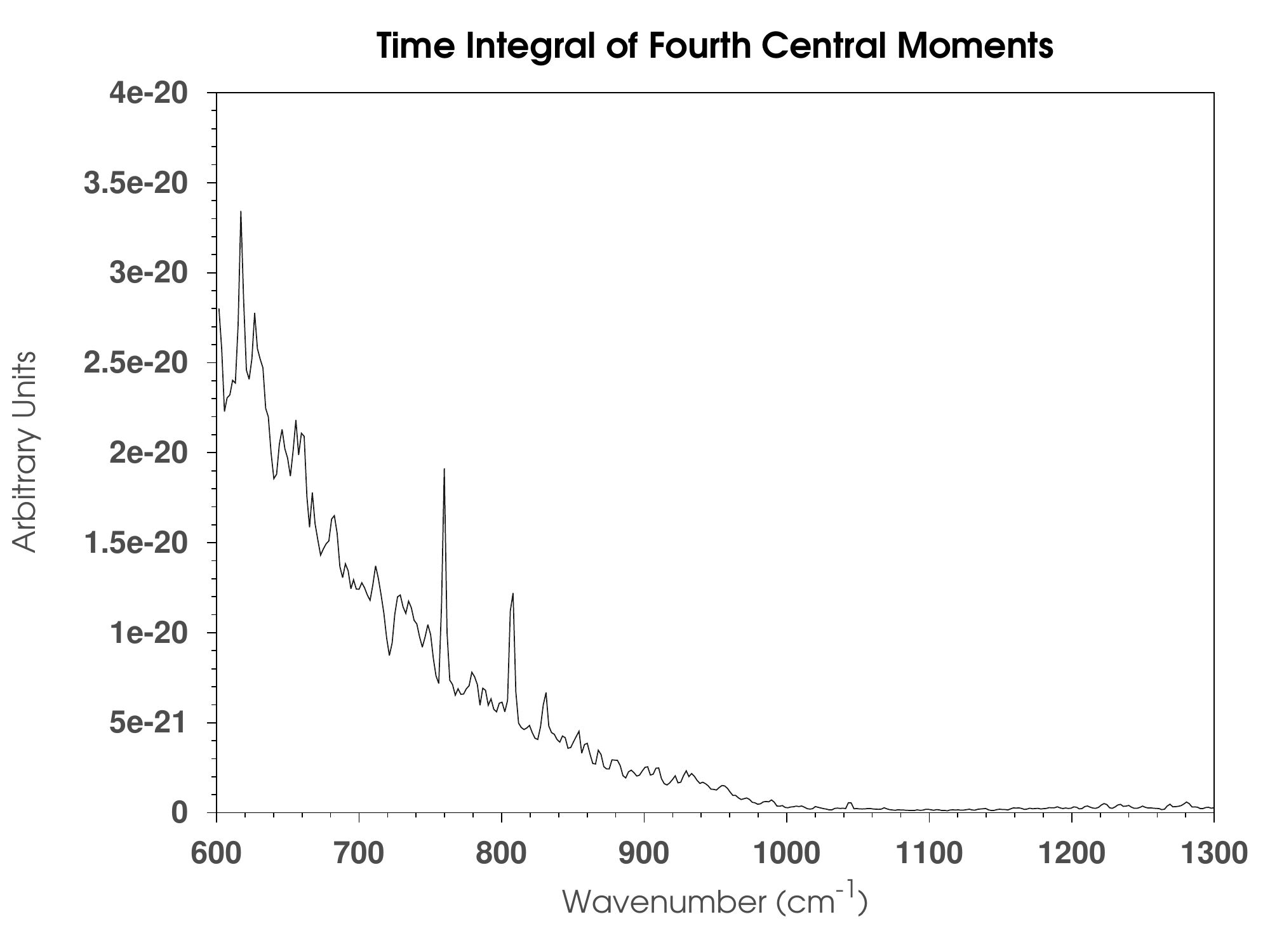}
\caption{The time average of the kurtosis of each spectral band, where the kurtosis is evaluated within a sliding window from sample $t$ to $t+30$.}
\end{figure}

The freqency average of the kurtosis across the spectrum can be seen in Figure 18, and the time average of each spectral band is shown in Figure 19.

\subsection{The $L_{\infty}$ norm}

Considering $\lim_{n \to \infty}$ of $L_n(X)$, as discussed earlier, we may calculate the upper limit of the p-norms, the $L_{\infty}$ norm.  For the example of the stressed rock, the result is shown in Figure 20:

\begin{figure}[H]
\centering
\includegraphics[width=9cm]{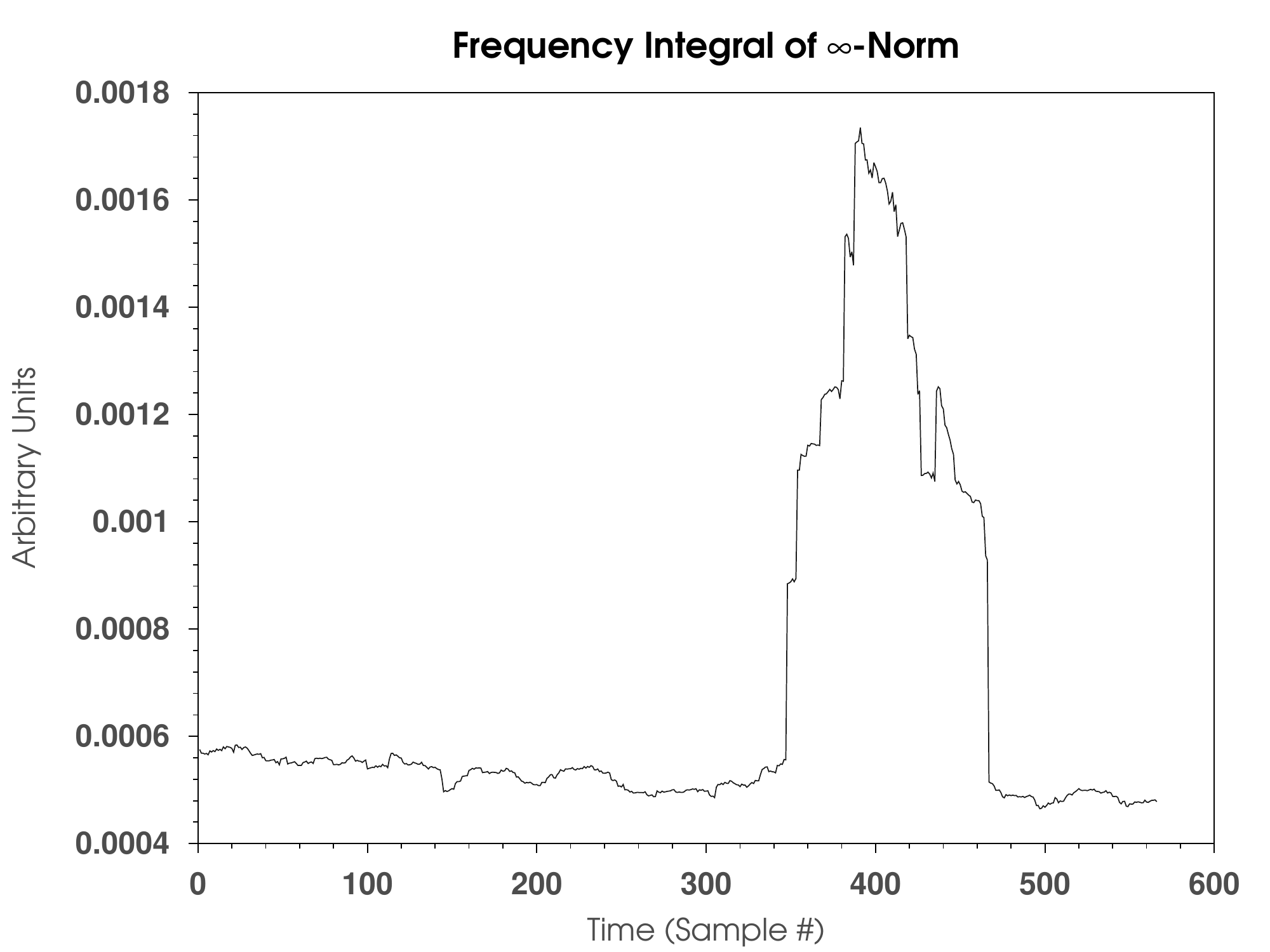}
\caption{The $L_{\infty}$ norm shows sharp changes in response to non-stationary behavior.}
\end{figure}

The $L_{\infty}$ norm changes rapidly in response to non-stationary behavior as the rock begins to fracture.  Additionally, relatively abrupt changes may be seen at several other points, and these points also seem to correspond to changes in the oscillatory behavior associated with fractional moments, as shown below.  The $L_{\infty}$ norm becomes more variable around $t=100$, at the time that force is initially applied to the rock.  Spectral features such as emission lines are relatively apparent in the time average of the $L_{\infty}$ norm.
\begin{figure}[H]
\centering
\includegraphics[width=9cm]{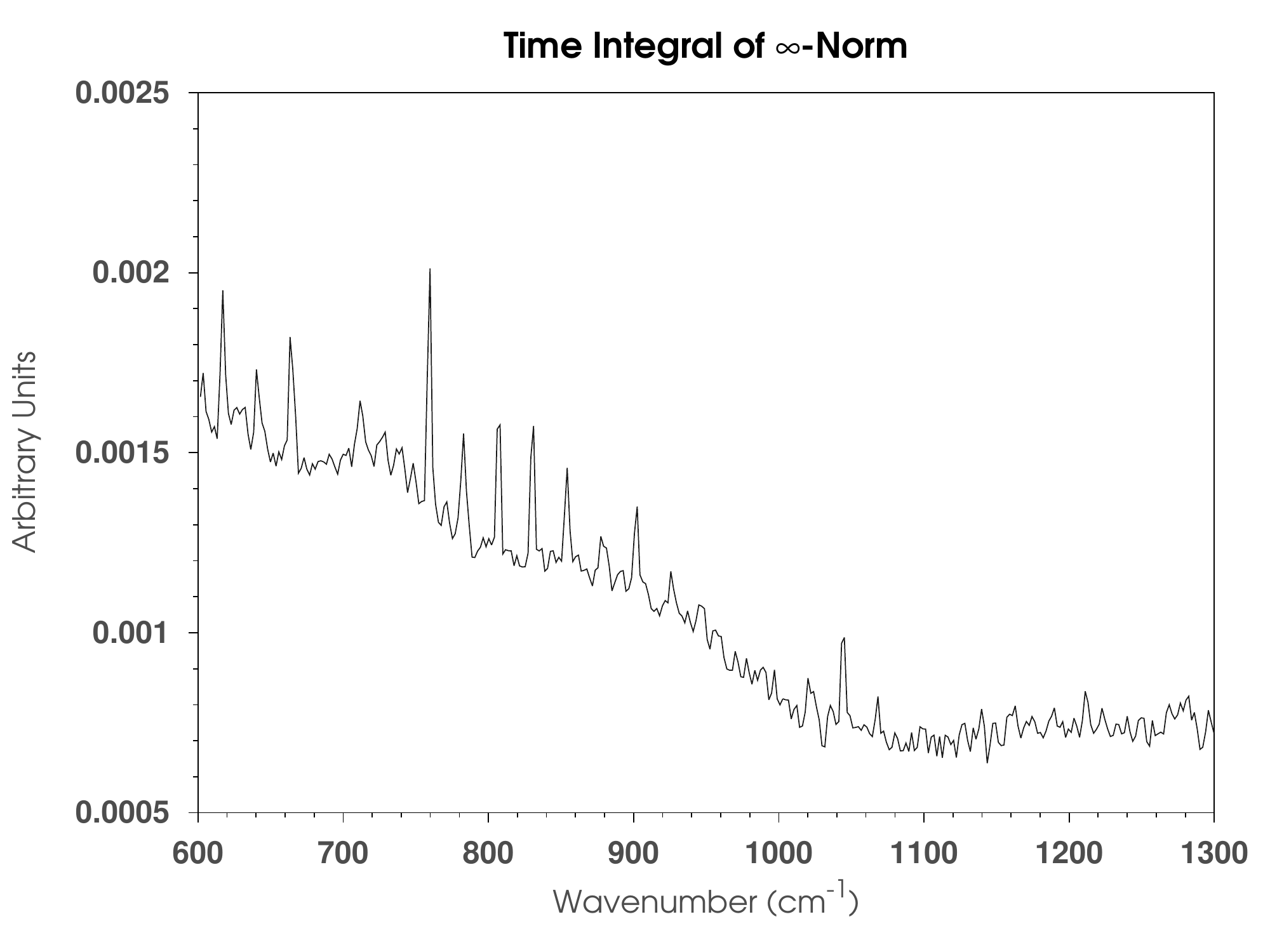}
\caption{The time-average of the $L_{\infty}$ norm shows prominent spectral lines similar to the time-average of the variance.}
\end{figure}

\subsection{The 50th moment}

\begin{figure}[H]
\centering
\includegraphics[width=9cm]{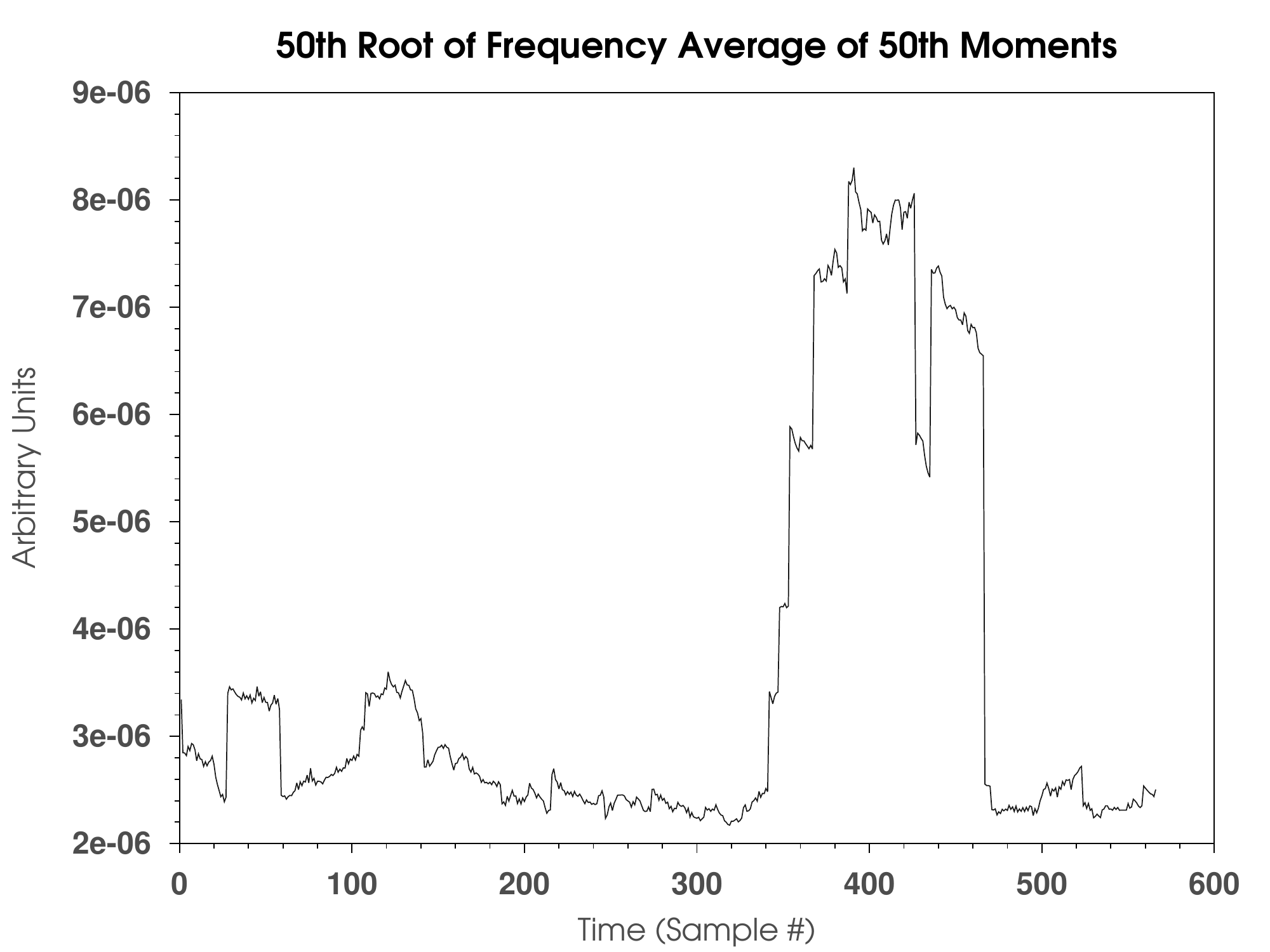}
\caption{High moments, such as the 50th, change sharply in response to non-stationary behavior.  The result is similar to that seen from the $\infty$-norm.  By taking the 50th root of the average 50th moment, we obtain units that are are linear in energy.  Again, a 30-sample sliding window was used.}
\end{figure}

Like the $L_{\infty}$ norm, high moments such as the 50th change rapidly in response to non-stationary behavior.  Unlike the $L_{\infty}$ norm, however, fewer spectral lines are apparent.

\begin{figure}[H]
\centering
\includegraphics[width=9cm]{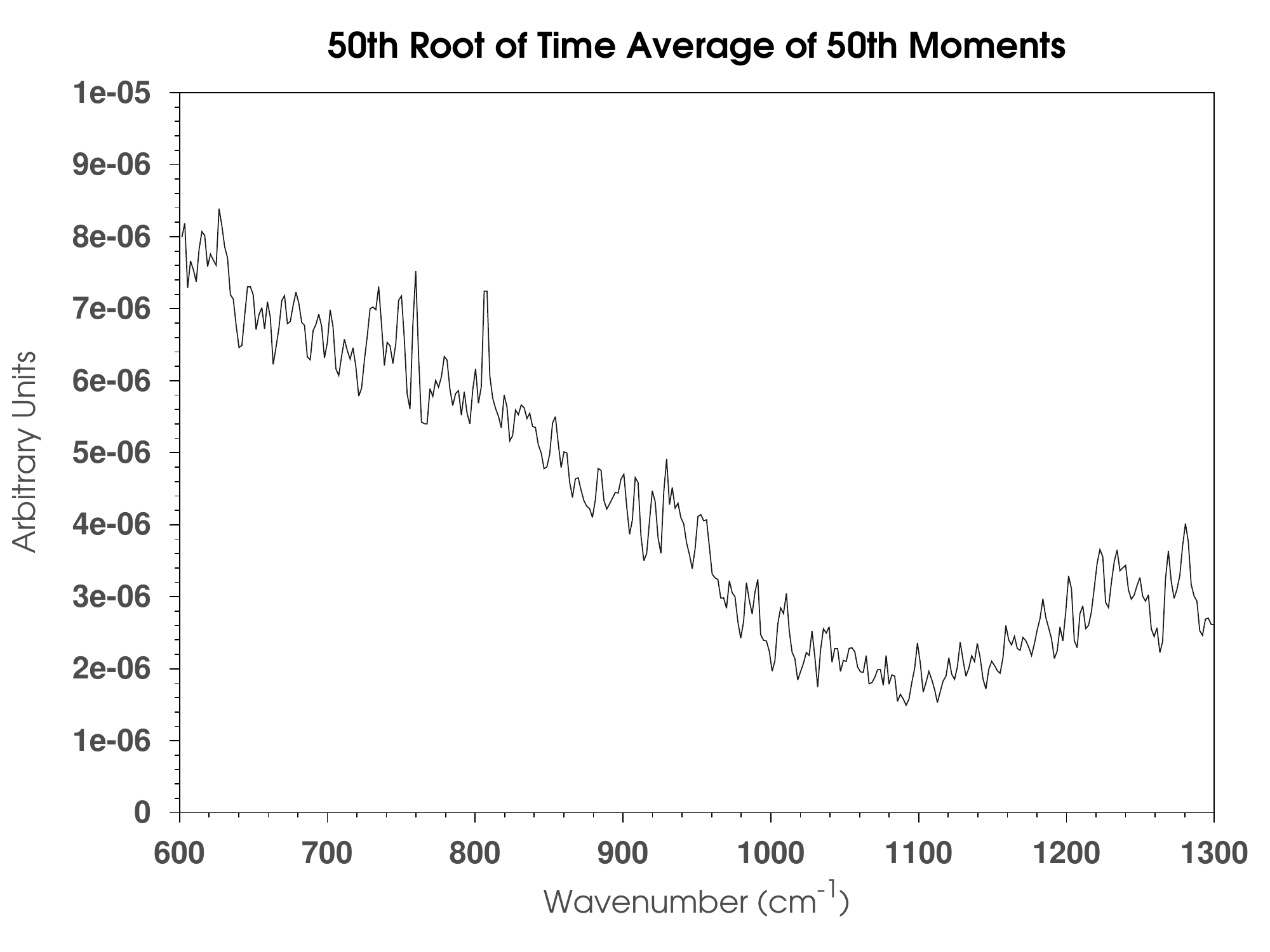}
\caption{Compared to $L_{\infty}$, fewer spectral lines are apparent from the 50th moment.  Again, the 50th root has been taken to maintain linearity in the y-axis units.}
\end{figure}

\subsection{A small fractional p-(quasi)norm}

Finally, we give an example of a fractional p-quasinorm.  As mentioned earlier, an oscillatory component is introduced at non-integer values of fractional moments and norms, and this oscillation is extremely sensitive to the distribution.  We calculate the real part of the 0.01-norm for the example of the stressed rock in a sliding window of width 30.

\begin{figure}[H]
\centering
\includegraphics[width=9cm]{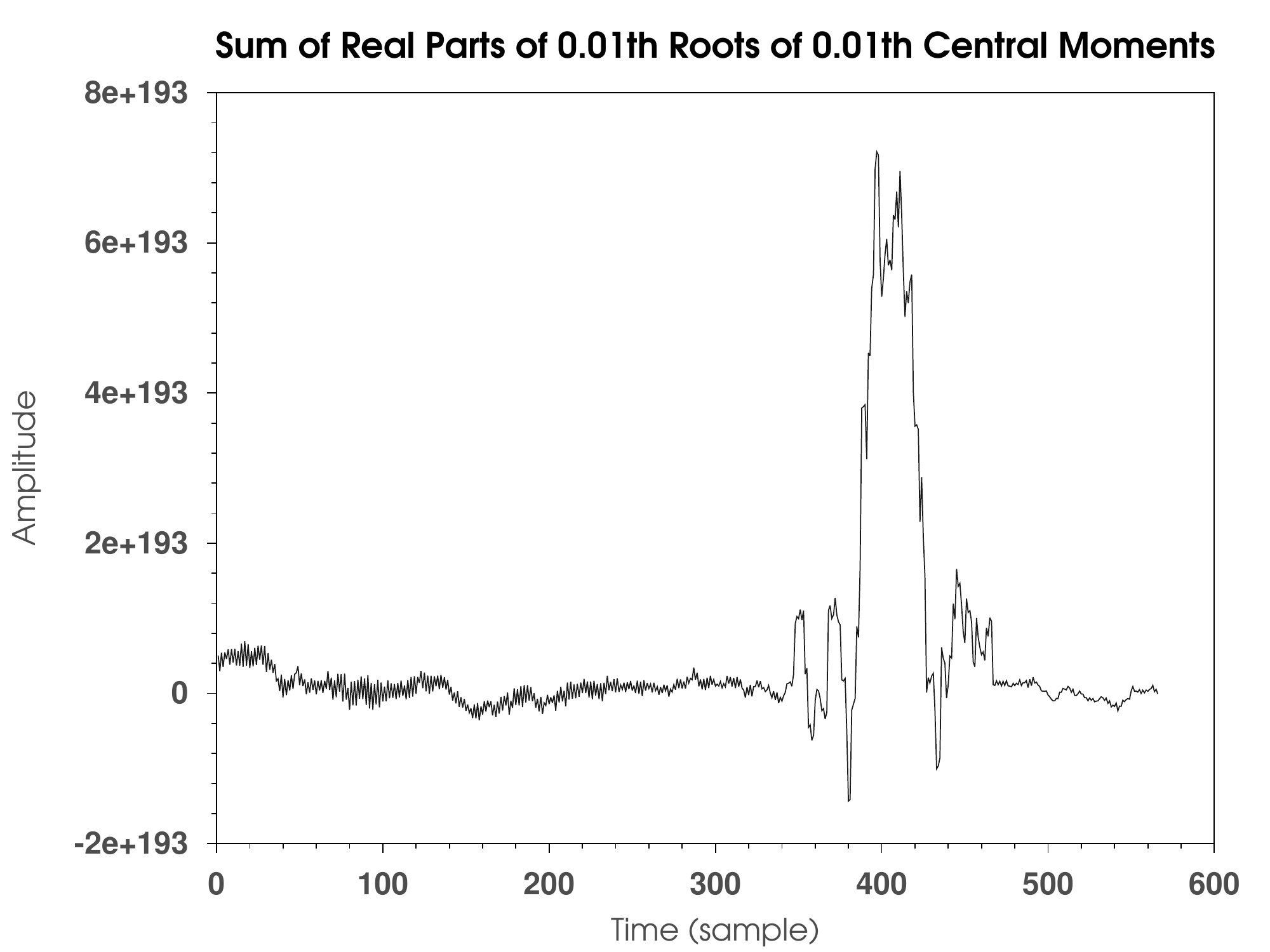}
\caption{The real part of the frequency average of the 0.01-norm introduces oscillatory components to the time series.  It can be seen that the amplitude of this oscillation changes, indicating a change in the statistical behavior of the time series.  This change becomes apparent at an earlier time that those seen in the nonfractional moments.}
\end{figure}

The 'spectra' obtained by averaging the 0.01-norm are somewhat more difficult to interpret.  Some lines are visible, but relatively large oscillations have been introduced, making spectral features difficult to resolve.

\begin{figure}[H]
\centering
\includegraphics[width=9cm]{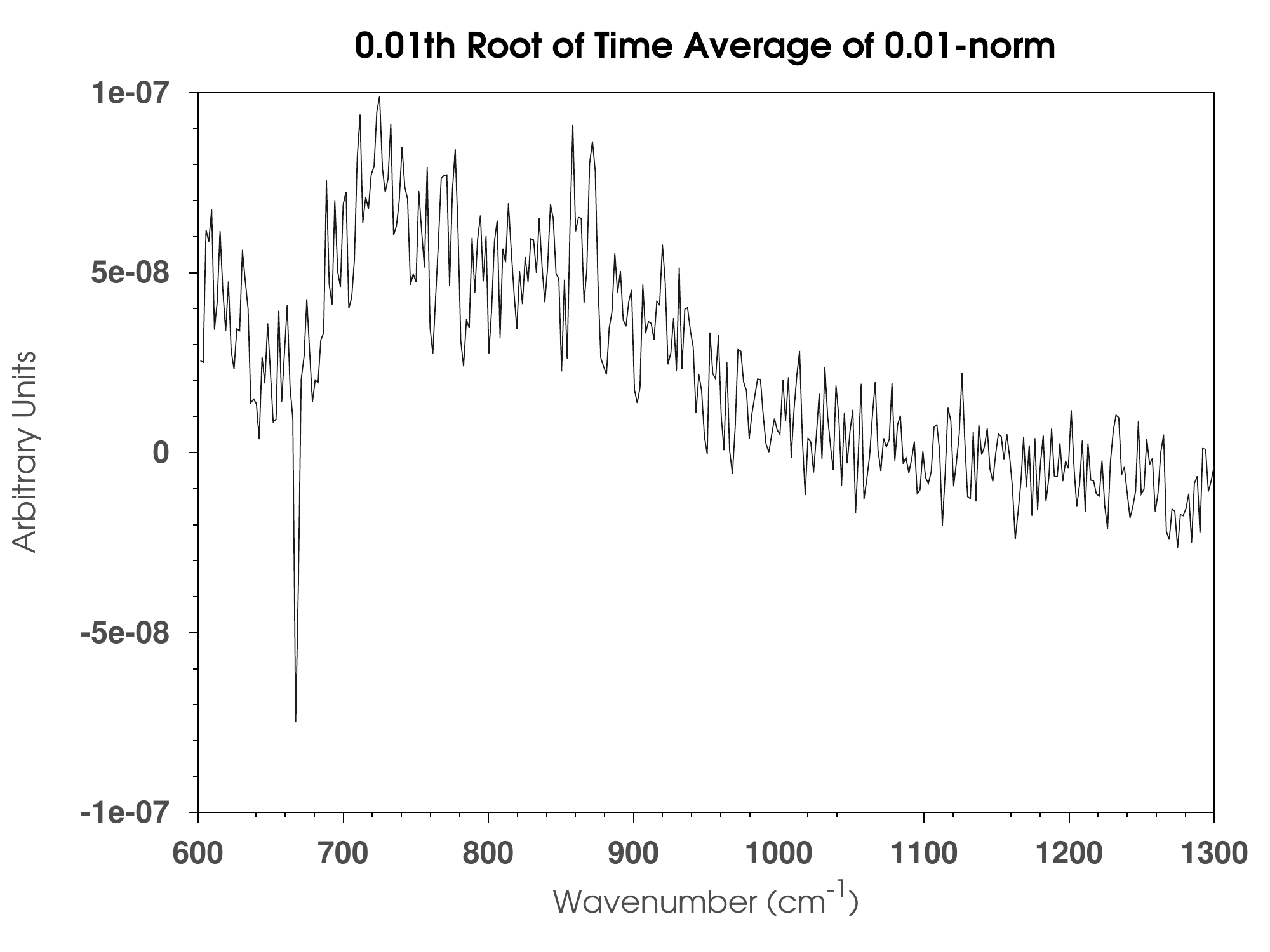}
\caption{The time average of $L_{0.01}$ is affected by complex oscillations, making spectral features less apparent.  A carbon dioxide line is still visible at 667cm$^{-1}$, the result otherwise bears resemblance to the time-average of the first prinicpal component, having a similar cutoff around 700cm$^{-1}$.}
\end{figure}

The oscillation induced by the fractional p-quasinorm is seen to have a much higher amplitude prior to fracture than after fracture.  Moreover, the change in the oscillatory behavior of the fractional moment is apparent at earlier times than in the amplitudes of the non-fractional moments (or from force/displacement of the hydraulic press).  This oscillation may be analyzed using a short-time Fourier transform.  The oscillation becomes consistently smaller in amplitude around $t=250$, well before the large changes that occur around $t=340$, and, somewhat astoundingly, well before any fracturing was apparent from mechanical measurements.  Fractional p-quasinorms can potentially identify very subtle changes in the stochastic behavior of a sequence.

\section{Conclusion}

Moments of spectral time series can identify complex and subtle behavior that would otherwise be obscured by noise.  While fully resolving spectral fluctuations about equilbrium is not possible with the relatively low sampling rates afforded by many spectrometers, the time-dependent statistical distribution of even severely undersampled spectral time series can reveal key information about the state of a system.  This time-dependent distribution can be parameterized by calculating the first few moments about the mean in sliding windows that pass over the time series of each spectral band.  This has the effect of emphasizing random and non-equilbrium behavior over equilibrium behavior.

Since a series of scans are normally averaged to produce a 'clean' spectrum, and the computational overhead associated with this method is minimal, the cost of implementing these techniques in modern hardware, e.g. FTIR spectrometers, is negligible.  An enhanced spectrum, encoding the statisical distribution of each spectral band, could easily be returned in the form of a collection of spectral moments or cumulants.  The first moment would simply be the averaged spectrum, the second moment the variance of the series of scans, and so on with higher moments, until all the useful information about the distribution of scans has been captured.  Usually, valuable information about the higher moments is unceremoniously discarded as useless noise.

This method has been shown to produce spectral lines that agree with standard references, and it has also been shown to produce much useful information about the state of one type of rock, Red Granite, subjected to a particular stimulus, namely, increased pressure.  Although the ability to identify subtle state changes in rock is interesting in itself, more study is necessary to determine what sorts of spectroscopy could benefit from this analysis, and under what conditions.  Moreover, additional work is needed to interpret the resulting statistics in the context of particular applications.  Nonetheless, the method seems capable of producing results that are comparable or superior to PCA, and with reduced computational overhead.  Early results are very encouraging that this analysis could enable new types of spectroscopic measurements in a wide variety of applications.

\begin{acknowledgements}
The author would like to acknowledge Dr. Friedemann Freund, who has designed experiments necessitating advances in spectroscopic analysis.  Portions of this work were funded by NASA Earth Surface and Interior Grant NNX12AL71G.
\end{acknowledgements}

\end{document}